\documentclass[a4paper,11pt]{article}

\usepackage{geometry}										% [showframe]
\usepackage[english]{babel}									% language
\usepackage[utf8]{inputenc}									% input text encoding
\usepackage[T1]{fontenc}									% output encoding

\usepackage{lmodern}										% Latin Modern

\usepackage{amssymb,amsfonts,amsmath,amsthm}
\usepackage{booktabs}
\usepackage{makecell}

\usepackage{algorithm}
\usepackage{algorithmic}

\usepackage{hyphenat}
\usepackage{caption}
\usepackage{subcaption}
\usepackage{enumitem}
\usepackage{graphics}
\graphicspath{{./JournalFigures/}}
\usepackage{mathtools}
\usepackage{etoolbox}
\usepackage{todonotes}

\newcommand{\Esp}[1]{{\mathbb E}\left[ #1 \right]}
\newcommand{\Espe}[2]{{\mathbb E}_{#1}\left[#2\right]}
\newcommand{\Var}[1]{{\rm Var}\left[ #1 \right]}

\newcommand{\ve}[1]{\boldsymbol{#1}}
\newcommand{\acc}[1]{\left\{#1\right\}}
\newcommand{\eqdef}{\stackrel{\text{def}}{=}}

\DeclarePairedDelimiter\abs{\lvert}{\rvert}
\DeclarePairedDelimiter\norm{\lVert}{\rVert}
\makeatletter
\let\oldabs\abs
\def\abs{\@ifstar{\oldabs}{\oldabs*}}
\let\oldnorm\norm
\def\norm{\@ifstar{\oldnorm}{\oldnorm*}}
\makeatother

\newcommand{\ca}{{\mathcal A}}

\newcommand{\cd}{{\mathcal D}}

\newcommand{\cl}{{\mathcal L}}
\newcommand{\cm}{{\mathcal M}}
\newcommand{\cn}{{\mathcal N}}
\newcommand{\ct}{{\mathcal T}}

\newcommand{\cx}{{\mathcal X}}
\newcommand{\cy}{{\mathcal Y}}

\newcommand{\Nn}{{\mathbb N}}

\newcommand{\D}{\mathrm{d}}

\newlength{\HYDROsubWidth}	
\newlength{\HYDROfigHeight}	
\newlength{\HYDROmapHeight}	
\newlength{\HYDROfigHeightNew}
\usepackage{authblk}										% [auth-sc,affil-it], to be loaded after titling
\title{Seismic fragility analysis using stochastic polynomial chaos expansions}
\author[1]{Xujia Zhu \thanks{zhu@ibk.baug.ethz.ch}}
\author[2]{Marco Broccardo\thanks{marco.broccardo@unitn.it}}
\author[1]{Bruno Sudret\thanks{sudret@ethz.ch}}
\affil[1]{Chair of Risk, Safety and Uncertainty Quantification, ETH Z\"{u}rich, Switzerland}
\affil[2]{Department of Civil, Environmental and Mechanical Engineering, University of Trento, Italy}

\date{\today}

\usepackage{microtype}										% typographical refinements

\usepackage{indentfirst}									% indents first paragraph in new section

\usepackage{caption}
\usepackage{subcaption}

\usepackage[numbers,sort&compress]{natbib}							% number ranges are compressed and hyphenated
\usepackage{hyperref}
\hypersetup{
pdftitle = {Seismic fragility analysis using stochastic polynomial chaos expansions},
pdfauthor = {Xujia Zhu, Marco Broccardo, Bruno Sudret},
pdfsubject = {Manuscript submitted to Probabilistic Engineering Mechanics},
pdfkeywords = {earthquake engineering, fragility functions, stochastic simulator, polynomial chaos expansion},
colorlinks = false,										% false: boxed links; true: colored links
}

\usepackage[capitalize]{cleveref}								% Eq., Fig., Table, Section, has to be loaded after hyperref
\creflabelformat{equation}{#2(#1)#3}								% hyperlinks covers the parenthesis around equation
\crefrangelabelformat{equation}{#3(#1)#4 to #5(#2)#6}						% multiple references
\labelcrefformat{subequation}{#2(#1)#3}								% the same for subequations
\labelcrefrangeformat{subequation}{#3(#1)#4 to #5(#2)#6}					% mutiple references

\begin{document}

\maketitle

\begin{abstract}
Within the performance-based earthquake engineering (PBEE) framework, the fragility model plays a pivotal role. Such a model represents the probability that the engineering demand parameter (EDP) exceeds a certain safety threshold given a set of selected intensity measures (IMs) that characterize the earthquake load. The-state-of-the art methods for fragility computation rely on full non-linear time-history analyses. Within this perimeter, there are two main approaches: the first relies on the selection and scaling of recorded ground motions; the second, based on random vibration theory, characterizes the seismic input with a parametric stochastic ground motion model (SGMM). The latter case has the great advantage that the problem of seismic risk analysis is framed as a forward uncertainty quantification problem. However, running classical full-scale Monte Carlo simulations is intractable because of the prohibitive computational cost of typical finite element models. Therefore, it is of great interest to define fragility models that link an EDP of interest with the SGMM parameters --- which are regarded as IMs in this context. The computation of such fragility models is a challenge on its own and, despite a few recent studies, there is still an important research gap in this domain. This comes with no surprise as classical surrogate modeling techniques cannot be applied due to the stochastic nature of SGMM. This study tackles this computational challenge by using \emph{stochastic polynomial chaos expansions} to represent the statistical dependence of EDP on IMs. More precisely, this surrogate model estimates the full conditional probability distribution of EDP conditioned on IMs. We compare the proposed approach with some state-of-the-art methods in two case studies. The numerical results show that the new method prevails over its competitors in estimating both the conditional distribution and the fragility functions. 
\end{abstract}

\section{Introduction}
\noindent The PEER\footnote{Pacific Earthquake Engineering Center} performance-based earthquake engineering (PBEE) framework introduced two decades ago \cite{Cornell2000} represents the state-of-the-art approach to seismic risk assessment. The framework builds on the total probability theorem by convolving the output of probabilistic seismic hazard analysis (PSHA, \cite{Cornell1968}) with fragility, damage, and loss models. The output of the PSHA analysis is the so-named hazard curves, which are rates of occurrence of a given \emph{intensity measure} (IM, e.g., peak ground acceleration, spectral acceleration, etc.) or a vector of IMs. The damage of a structure is typically characterized by the \emph{engineering demand parameter} (EDP) which represents the structural response (e.g., the maximum interstory drift for a multistory building, the maximum base shear, etc.). 

A critical component of the framework is represented by the statistical relationship between IMs and EDP. This relationship, named fragility model, is a function of the IMs and computes the EDP exceeding probability (e.g., EDP exceeds a certain threshold) conditioned on the corresponding value of IMs.  As an important part of PBEE, fragility models have become a rich field of research with two major lines of investigation. The first line is based on the selection and scaling of recorded ground motions and (non-)linear time history analysis. An incomplete list of studies following this line of research includes \cite{Vamvatsikos2002,Baker2006,Luco2007,Kiani2015}. 

The second line of research builds on stochastic ground motion models (SGMM) \cite{Rezaeian2008,Rezaeian2010}, and (non-)linear time history analysis. An SGMM typically combines a set of engineering-meaningful parameters, referred to as SGMM parameters in the sequel, with a set of hidden aleatory variables (e.g., white noise) to generate synthetic ground motions. The available records are considered as realizations of the SGMM and used to calibrate the SGMM parameters. The latter are modeled as random variables to account for epistemic uncertainties due to limited data. In this setting, the SGMM parameters are statistically related to the earthquake and site characteristics (e.g., magnitude, faulting mechanism, source-to-site distance, and the site shear-wave velocity) via predictive equations. In essence, these are classical ground motion predictive equations (GMPEs, \cite{Cornell1968}) with the IMs being the SGMM parameters. 

Following this line, a fragility model becomes the statistical relationship between the SGMM parameters and the EDP. These models, when developed, allow for a rapid seismic risk assessment by computing directly or via (inexpensive) simulations of the convolutions of the PEER-PBEE framework. Within this perimeter, therefore, the development of efficient algorithms for fragility computation is paramount. While several studies use an SGMM for seismic risk assessment (an incomplete list includes \cite{Taflanidis2009,Gidaris2015,Mai2017,Smerzini2018,Ghosh2020,Ghosh2021}), to the best of our knowledge, fragility models as a function of the SGMM parameters have been explicitly introduced only recently, \cite{Abbiati2021}. 

In this context, however, there is a research gap in the development of efficient algorithms that allow a feasible computation of these special fragility models. This paper aims to fill this gap by using the stochastic polynomial chaos expansion (SPCE) \cite{ZhuIJUQ2022}, which we show to be the most computationally efficient option up to date. As such, this paper focuses only on the fragility model computation without employing the full seismic risk analysis.

A great advantage of the simulation-based approach is that the problem of seismic risk analysis can be framed as a forward uncertainty quantification problem \cite{Abbiati2021}. In fact, by combining the SGMM with the dynamical analysis of structures, one obtains a simulator that maps a set of ground motion parameters to the associated EDP. More specifically, this is a \emph{stochastic simulator} \cite{ZhuIJUQ2020,Abbiati2021}, i.e., several runs with the same ground motion parameters produce different values of the EDP, due to the aleatory hidden variables in the generation of ground motions. Therefore, one can run multiple simulations for given values of IMs without introducing bias. Moreover, this allows for coupling the seismic hazard model and the fragility function without going through intermediate variables. 

When working with ground motion parameters, replication-based methods have been proposed so far in the literature \cite{Gidaris2015,Abbiati2021}. 
In this framework, one fixes the SGMM parameters, and the hazard model produces a set of consistent earthquake loads for dynamical analysis of the structure. This procedure is called \emph{replication}, as we evaluate repeatedly the simulator for the same values of the input. The associated EDP values are realizations of the structural response conditioned on the given SGMM parameters. Therefore, they can be used to estimate the underlying conditional distribution. This procedure is repeated for different SGMM parameters, and the fragility function can be estimated from the conditional distribution. Because many replications (e.g., $100$) are necessary to characterize the conditional distribution, this approach requires a large number of model runs (as shown in \cite{Abbiati2021}). 

To alleviate the computational cost, in this paper, we explore the methods that \textit{do not} rely on replications \cite{Cornell2002,Shinozuka2000,Mai2017}. Since the SGMM parameters are vector-valued IMs, some methods developed for fragility analysis with a single IM can be extended and applied. Cornell et al. \cite{Cornell2002} proposed the so-called cloud analysis which is a linear model in the log-scale with a homoscedastic Gaussian noise. This parametric model relies on rather restrictive assumptions (log-linearity and homoscedasticity). 

Alternatively, fragility models can be computed in a classification framework \cite{Shinozuka2000, Baker2015}. This method only works with binary damage variables (whether the structure fails or not) and does not make use of the precise value of the EDP, which leads to a certain loss of information. More recently,  nonparametric models, namely kernel smoothing, have been proposed in the literature \cite{Noh2015,Mai2017}. However, it is well-known that nonparametric models suffer from the curse of dimensionality \cite{Tsybakov2009}: the model accuracy decreases drastically with increasing input dimensionality (in our case, the number of IMs). 

In this paper, to better balance the model flexibility and the limited number of simulations, we propose applying the newly developed \emph{stochastic polynomial chaos expansion} (SPCE) technique \cite{ZhuIJUQ2022}. This model introduces an artificial latent variable and a noise variable to represent the random nature of the stochastic simulation. More precisely, it expresses the EDP as a function of the IMs and the latent variable plus the additive noise. Therefore, this model can tackle a full representation of conditional distributions. It follows that natural byproducts of the analysis are the \textit{classical} fragility models. In fact, one can naturally develop statistical relations between classical IMs and the selected EDP. In this case, the classical IMs are available as statistics of the synthetic ground motions\footnote{In this case, one has to verify that the rate of exceedance of the classical IMs emerging from the SGMM is compatible with the ones derived by PSHA analysis \cite{Rezaeian2010}}, and the fragility models can be used in the original PEER-PBEE framework directly.

The paper is organized as follows. In \Cref{sec:FA}, we outline the stochastic simulator approach; then, we recap the extension of classical methods developed to multiple intensity measures. In \Cref{sec:SPCE}, we summarize the main ingredients of the stochastic polynomial chaos expansion. In \Cref{sec:examples}, we use a synthetic ground motion model and two computational examples to illustrate the performance of the proposed method. Finally, we conclude with the main finding of the study and give an outlook for future research  in \Cref{sec:conclusion}.

\section{Stochastic simulator approach for fragility analysis}
\label{sec:FA}
\subsection{The stochastic simulator approach}
\noindent
This paper follows the line of research that uses an SGMM  to characterize seismic excitation. Using the representation introduced in Abbiati et al. \cite{Abbiati2021}, the stochastic ground motion can be expressed as follows
\begin{equation}\label{eq:SGMM_ex}
	A(t) = \mathcal{M}_a(t,\ve{\Xi}|\ve{X}), 
\end{equation}
where $\mathcal{M}_a$ represents the synthesis formula of a parametric SGMM, $\ve{\Xi}$ is a Gaussian vector (with i.i.d. standard normal random variables) representing the aleatory variability of the process, and $\ve{X}$ is a random vector collecting the parameters of the model and the associated epistemic variability. The SGMM parameters are selected to be engineering meaningful (\cite{Rezaeian2008, Broccardo2017}); therefore, in this framework, $\ve{X}$ can be regarded as a vector of IMs. In the PEER-PBEE framework, $\ve{X}$ is statistically related to the earthquake and site characteristics via predictive equations. However, this study focuses only on the fragility model computation and, therefore, for simplicity, we use a marginal joint probability distribution of $\ve{X}$ fitted to a specific seismic catalog (see \Cref{sec:sgm} for further details). 

Let $Y$ denote the EDP (e.g., maximum interstory drift) of a structural system of interest computed as $Y = \mathcal{M}_d(A(t)|\ve x_d)$, where $\mathcal{M}_d$ is an expensive-to-evaluate deterministic solver\footnote{In \cite{Abbiati2021}, the solver is also assumed to be stochastic to accommodate random fields. In this paper, we choose the more restrictive deterministic solver as it is the most typical case in earthquake engineering} with $\ve{x}_d$ being a set of deterministic parameters (e.g., a finite element model with deterministic masses, damping, and constitutive models). It follows that $Y$ can be expressed as
\begin{equation}\label{eq:stosim}
	Y = \mathcal{M}_d(\mathcal{M}_a(t,\ve{\Xi}|\ve{X})|\ve{x}_d)=\mathcal{M}_s(\ve{\Xi}|\ve{X}),
\end{equation}
where $\mathcal{M}_s \eqdef \mathcal{M}_d \circ \mathcal{M}_a$ is a stochastic simulator since for $\ve{X}=\ve{x}$ the response $Y$ is still stochastic (due to the aleatory variability encoded in $\ve{\Xi}$). Provided with this framework, the objective of this study is to use a stochastic surrogate model, namely the SPCE, to develop fragility models. 

\noindent

\subsection{Fragility analysis}
\label{sec:CFA}
\noindent In PBEE, seismic loads are typically characterized by a selected set of IMs. An incomplete list of conventional IMs includes peak ground acceleration, spectral acceleration, peak ground velocity, and Arias intensity \cite{mackie2003seismic}. In general, an IM can represent any ``optimal'' feature of the seismic load. According to \cite{mackie2003seismic}, optimal is defined as being practical, sufficient, effective, and efficient (see \cite{mackie2003seismic} for further details). To improve the power of the prediction and reduce the variability among ground motions, one can combine several IMs for fragility analysis \cite{Baker2005,Seyedi2010,Modica2014}. 

In the SGMM context, a natural choice for the IMs is the set of SGMM parameters. This allows applying directly the PBEE-PEER framework by convolving the predictive equations (which extend the classical GMPEs) with these fragility models based on the SGMM parameters \cite{Abbiati2021}. In this study, we pursue this philosophy by proposing SPCE as a computational method that outperforms the current state of the art. In particular, this section first introduces the general concept of fragility models; second, it reviews a series of computational methods which can be used directly in this context and that we will use to compare the proposed SPCE approach.

The structural performance is usually defined by the event that the EDP exceeds a certain threshold $\delta_0$, which represents a predefined damage level. A fragility model expresses the exceeding probability as a function of IMs, that is,
\begin{equation}\label{eq:fragility}
		p_f(\ve{x}) = \mathbb{P}\left(Y>\delta_0\mid \ve{X}=\ve{x}\right) = 1 - F_{Y\mid\ve{X}}(\delta_0\mid \ve{x}).
\end{equation}
\noindent
Using the distribution characterizing the SGMM parameters, we generate $N$ samples grouped into $\cx=\acc{\ve{x}^{(1)},\ldots,\ve{x}^{(N)}}$. Unlike Gidaris et al. \cite{Gidaris2015} and Abbiati et al. \cite{Abbiati2021}, where $\mathcal{O}(10^2)$ replications are used, we do not consider replications in this paper to drastically reduce the overall number of simulations. This is feasible because of the features of the SPCE approach described in \Cref{sec:SPCE}. Therefore, for each set of the ground motions parameters $\ve{x}^{(i)}$, we generate one synthetic ground motion and then compute the associated EDP $y^{(i)}$ which is collected in $\cy=\acc{y^{(1)},\ldots,y^{(N)}}$.

In the sequel, we introduce a series of classical fragility model computation methods, which can be used directly in this context. Moreover, we use these benchmark methods to compare the proposed SPCE approach. 

One of the most popular methods for fragility analysis is the linear model \cite{Cornell2002,Modica2014} (i.e., the so-called cloud analysis), where the logarithm of EDP is expressed as a linear function of the logarithm of the IMs with an independent additive Gaussian noise, i.e., 
\begin{equation}\label{eq:lm}
	\begin{split}
		\log (Y) &= \sum_{i=1}^{M}  \beta_0 + \sum_{j=1} \beta_j\log(x_j)  + e, 
	\end{split}
\end{equation}
where $e \sim \cn(0,\sigma^2)$. The model parameters $\ve{\beta}$ and $\sigma$ can be estimated using standard ordinary least-squares. \Cref{eq:lm} gives directly the conditional probability density function (PDF), and the fragility function is calculated as
\begin{equation}\label{eq:LM}
	p_f(\ve{x}) = 1-\Phi\left(\frac{\delta_0-\beta_0 - \sum\limits^{M}_{j=1} \beta_j\ln(x_j)}{\sigma}\right),
\end{equation}
where $\Phi$ is the cumulative distribution function (CDF) of the standard normal distribution.

Probit regression is another classical method used to estimate directly fragility functions \cite{Shinozuka2000, Baker2015}. In this context, fragility models are interpreted as a soft classifier. In the earthquake engineering community, the CDF of a lognormal distribution is typically selected as classifier. Although this method is usually used for a single IM, it can be extended directly to the case of multiple IMs, that is,
\begin{equation}\label{eq:probit}
	p_f(\ve{x}) = \Phi\left( \beta_0 + \sum_{j=1} \beta_j\ln(x_j) \right).
\end{equation}
The model parameters $\ve{\beta}$ are estimated by maximum likelihood estimation. In this classification framework, the threshed $\delta_0$ is used to directly classify the samples of the outcomes (e.g., $\{\text{not fail}\} \eqdef \{ EDP<\delta_0\}$, $\{\text{fail}\} \eqdef \{ EDP\ge\delta_0\}$), and the precise value of the EDP is ignored. Therefore, $\delta_0$ is a property of the classifier; in other words, when the value of $\delta_0$ varies, it is necessary to build a new model. 

In recent years, nonparametric methods for fragility model computations have gained momentum \cite{Noh2015,Mai2017}, given their inherent flexibility. Recall the definition of the conditional distribution
\begin{equation}\label{eq:nonparam}
	f_{Y\mid\ve{X}}(y\mid\ve{x}) = \frac{f_{Y,\ve{X}}(y,\ve{x})}{f_{\ve{X}}(\ve{x})}.
\end{equation}
Without introducing restrictive assumptions, the distributions $f_{Y,\ve{X}}$ and $f_{\ve{X}}$ can be estimated using nonparametric estimators, namely kernel smoothing, which then provides an estimate of the conditional distribution. In this approach, the bandwidths are hyper-parameters to be defined. Noh et al. \cite{Noh2015} proposed selecting the bandwidths by engineering judgments and prior information. Mai and Sudret \cite{Mai2017} applied the method developed in \cite{Duong2005} to estimate separately $f_{Y,\ve{X}}$ and $f_{\ve{X}}$. However, this does not yield a valid conditional distribution (the integral over $y$ is unequal to $1$). In this paper, we consider a more advanced nonparametric method developed by Li et al. \cite{Li2013} that is typically designed for estimating the conditional CDF, as the latter is directly related to the exceeding probability. Following Mai and Sudret \cite{Mai2017}, the kernel estimator is applied to the logarithmic transform of the data to guarantee the positiveness of the EDP and the IMs.

\section{Stochastic polynomial chaos expansion}
\label{sec:SPCE}
\noindent
The methods reviewed in the previous section have their limitations: the linear model relies on very restrictive assumptions, the probit model does not make full use of the information, and the kernel estimator suffers from the curse of dimensionality \cite{Tsybakov2009}. To achieve better accuracy with a limited number of simulations, we propose using the stochastic polynomial chaos expansion (SPCE) approach recently proposed in Zhu and Sudret \cite{ZhuIJUQ2022} to estimate the probability distribution of the EDP, $Y$, conditioned on the IMs, $\ve{X} = \ve{x}$. The conditional random variable is denoted by $Y_{\ve{x}}$. In this section, we recap the principle of the standard polynomial chaos expansion (PCE) and its extension to SPCE.

PCE is a surrogate model that has been widely applied to emulate deterministic simulators in the context of uncertainty quantification. Considering the uncertain input variables $\ve{X}$, this surrogate represents a deterministic model $\cm_d: \ve{x} \mapsto \cm_d(\ve{x})$ by a series of polynomial expansions, that is,
\begin{equation}\label{eq:PCE}
	\cm_d(\ve{X}) \approx \sum_{\ve{\alpha} \in \ca} c_{\ve{\alpha}} 
	\psi_{\ve{\alpha}}(\ve{X}),
\end{equation}
where $\psi_{\ve{\alpha}}$ is the basis function defined by the multi-index $\ve{\alpha}$, $c_{\ve{\alpha}}$ is the associated coefficient, and $\ca$ is the truncated set of multi-indices that define the basis functions used in the expansion. 

For $\ve{X}$ with independent components, the basis function is given by a product of univariate polynomials:
\begin{equation}\label{eq:PCE_basis}
	\psi_{\ve{\alpha}}(\ve{x}) = \prod_{j=1}^{M} \phi^{(j)}_{\alpha_j}(x_j),
\end{equation}
where $M$ is the dimension of $\ve{X}$, i.e., the number of input parameters, $\alpha_j$ is the polynomial degree in $x_j$, and $\acc{\phi^{(j)}_k:k\in\Nn}$ is the orthogonal polynomial basis with respect to the marginal distribution $f_{X_j}$, which satisfies
\begin{equation}\label{eq:unPCE_ortho}
	\Esp{\phi^{(j)}_{k}(X_j)\,\phi^{(j)}_{l}(X_j)}=
	\begin{cases}
		1 &\text{if } l=k\\
		0 &\text{otherwise}
	\end{cases}.
\end{equation}
For uniform, normal, gamma, and beta distributions, the associated univariate orthogonal polynomials are well known as Legendre, Hermite, Laguerre, and Jacobi polynomials \cite{Xiu2002}.

When $\ve{X}$ has dependent components, the tensor product in \cref{eq:PCE_basis} generally does not produce an orthogonal basis. To circumvent this problem, one common way is to transform $\ve{X}$ into an auxiliary vector $\ve{H} = \ct(\ve{X})$ with independent components (e.g., a standard normal vector) using the Nataf or Rosenblatt transform \cite{Torre2019}. The polynomial basis is then defined with respect to the auxiliary variables
\begin{equation}\label{eq:PCE_trans}
	\psi_{\ve{\alpha}}(\ve{x}) = \prod_{j=1}^{M} \phi^{(j)}_{\alpha_j}(h_j).
\end{equation}
where $\ve{h} = \ct(\ve{x})$, and $\acc{\phi^{(j)}_k:k\in\Nn}$ is defined by the marginal distribution of $H_j$.

Let us introduce now the stochastic extension of PCE. \Cref{eq:PCE} is a deterministic function of the input variables $\ve{x}$. To represent the stochastic behavior in the earthquake simulation, we include an artificial latent variable $Z$ in the expansion and an additive noise variable $\epsilon$ which results in the SPCE \cite{ZhuIJUQ2022}:
\begin{equation}\label{eq:SPCE}
	\log \left(Y_{\ve{x}}\right) \stackrel{\rm d}{\approx} \log \left(\tilde{Y}_{\ve{x}}\right) = \sum_{\ve{\alpha} \in \ca} c_{\ve{\alpha}} 
	\psi_{\ve{\alpha}}\left(\ve{x},Z\right) + \epsilon,
\end{equation}
where the expansion is expressed on the logarithmic transform of $Y_{\ve{x}}$ to ensure the EDP is positive (this transform is also applied by Gidaris et al. \cite{Gidaris2015}). The noise variable $\epsilon$ is a centered Gaussian random variable with standard deviation $\sigma$, i.e., $\epsilon \sim \cn(0,\sigma^2)$. 

Here, we aim at approximating the distribution of the EDP $Y_{\ve{x}}$ for any $\ve{x}$. As a result, we use the notation $\stackrel{\rm d}{\approx}$ to denote \emph{approximation in distribution}. The artificial latent variable $Z$ in \cref{eq:SPCE} is only introduced to reproduce the stochasticity, and it is \emph{not} related to the high-dimensional hidden random vector $\ve{\Xi}$ in the stochastic ground motion model of \cref{eq:stosim}. In this paper, we select a standard Gaussian latent variable$Z \sim \cn(0,1)$. With this choice, if only linear terms are considered in \cref{eq:SPCE}, the SPCE is equivalent to the linear model in \cref{eq:lm}.

To build such a model, we need to determine the coefficients $\ve{c}$ of the expansion and the standard deviation $\sigma$ of the noise term. For a data point $(\ve{x},y)$ the conditional likelihood can be expressed as (see details in Zhu and Sudret \cite{ZhuIJUQ2022})
\begin{equation}\label{eq:likelihood}
	l(\ve{c},\sigma;\ve{x},y) = \frac{1}{y}\int_{\cd_{Z}} \frac{1}{\sqrt{2\pi}\sigma} 
	\exp\left(-\frac{\left(\log(y) - \sum_{\ve{\alpha} \in \ca} c_{\ve{\alpha}} 	
		\psi_{\ve{\alpha}}(\ve{x},z)\right)^2}{2\sigma^2}\right) f_{Z}(z) \D z.
\end{equation} 
In practice, we can apply the Gaussian quadrature \cite{Golub1969} with respect to the weight function $f_{Z}$ to efficiently evaluate the one-dimensional integral, that is
\begin{equation}\label{eq:quadrature}
	\begin{split}
		l(\ve{c},\sigma;\ve{x},y) &\approx \tilde{l}(\ve{c},\sigma;\ve{x},y)\\
		&= \frac{1}{y}\sum_{j=1}^{N_Q} \frac{1}{\sqrt{2\pi}\sigma}\exp\left(-\frac{\left(\log(y) - \sum_{\ve{\alpha} \in \ca} c_{\ve{\alpha}}
			\psi_{\ve{\alpha}}(\ve{x},z_j)\right)^2}{2\sigma^2}\right) w_j ,
	\end{split}
\end{equation}
where $N_Q$ is the number of integration points, $z_j$ is the $j$-th integration point, and $w_j$ is the associated weight. Based on \cref{eq:quadrature} and the available data $(\cx,\ve{y})$, we calibrate the coefficients $\ve{c}$ by maximum likelihood estimation (MLE)
\begin{equation}\label{eq:MLE}
	\hat{\ve{c}} = \arg\max_{\ve{c}} \; \sum_{i}^{N}
	\log\left(\tilde{l}\left(\ve{c},\sigma;\ve{x}^{(i)},y^{(i)}\right)\right).
\end{equation}

The standard deviation $\sigma$ cannot be fitted jointly with $\ve{c}$ because the likelihood in \cref{eq:likelihood} is unbounded for $\sigma=0$ (see \cite{ZhuIJUQ2022} for a detailed discussion). Therefore, $\sigma$ is a hyper-parameter, and we use cross-validation with the out-of-sample likelihood as the performance metric to select an optimal value for $\sigma$. In addition, the cross-validation score is also useful for determining an appropriate truncated set $\ca$.

After constructing the model, one can efficiently generate new samples of $\tilde{Y}_{\ve{x}}$ by fixing the value of $\ve{x}$ and sampling $(Z,\epsilon)$ to evaluate \cref{eq:SPCE}. Therefore, probabilistic quantities of $\tilde{Y}_{\ve{x}}$ (e.g., mean, variance, quantiles, and exceeding probabilities \cref{eq:fragility}) can be estimated by large-scale Monte Carlo simulations. Similarly, jointly sampling $(\ve{X},Z,\epsilon)$ produces samples of $\tilde{Y}$ which can be used to study the properties of the emulated EDP.

\section{Numerical examples}
\label{sec:examples}
\noindent In this section, we compare SPCE with the methods reviewed in \Cref{sec:CFA}, namely the linear model (LM) \cite{Cornell2002}, the kernel conditional distribution estimator (KCDE), and the classical classification-based fragility model (i.e., the probit model) \cite{Shinozuka2000}, on two numerical examples. For the KCDE, we apply the kernel estimator developed for conditional CDF estimation \cite{Li2013} which is available in the package \texttt{np} \cite{Hayfield2008} implemented in R. 
To quantitatively assess the performance, we report the convergence of the models for the estimation of the conditional distribution and the fragility function.

When comparing the distribution estimation, we consider only LM, SPCE, and KCDE, as the probit model directly estimates the fragility function without providing the conditional distribution. Since LM, SPCE, and KCDE are all applied to the logarithmic transform of the EDP, we examine the estimation accuracy of the conditional distribution of the transformed quantity. In this respect, we use the normalized Wasserstein distance \cite{ZhuIJUQ2022} as the error metric which reads
\begin{equation}\label{eq:Rlevel1} 
	\varepsilon = \frac{\Espe{\ve{X}}{d^2_{\rm	WS}\left(\log\left(Y_{\ve{X}}\right),\log\left(\tilde{Y}_{\ve{X}}\right)\right)}}{\Var{\log(Y)}},
\end{equation}
where $Y_{\ve{x}}$ is the EDP obtained from the stochastic simulation, $\tilde{Y}_{\ve{x}}$ is that of the surrogate model, and $d_{\rm WS}$ is the \emph{Wasserstein distance of order two} \cite{Villani2008} between two probability measures. For continuous random variables $Y_1$ and $Y_2$  with quantile functions (i.e., inverse CDF) $Q_1$ and $Q_2$, this distance can be computed by
\begin{equation}\label{eq:WS}
	d^2_{\rm WS}\left(Y_1,Y_2\right) = \norm{Q_1 - Q_2}^2_2 =  \int_{0}^{1}\left(Q_1(u) - Q_2(u)\right)^2\D u,
\end{equation}

For the fragility model in \cref{eq:fragility} which is a deterministic function of $\ve{x}$, we use the relative mean-squared error to assess the global approximation accuracy
\begin{equation}\label{eq:errorq}
	\varepsilon_{p} \eqdef \frac{\Esp{\left(p_f(\ve{X}) - 
		\tilde{p}_f(\ve{X})\right)^2}}{\Var{p_f(\ve{X})}},
\end{equation}
where $p_f$ is the fragility function of the simulator, and $\tilde{p}_f$ denotes that of the surrogate.

\subsection{Stochastic ground motion model}
\label{sec:sgm}
\noindent
This section briefly describes the simplified SGMM model used in our analysis. It is out of the scope of the current study to develop predictive equations that link the SGMM parameters to the earthquake site and source characteristics. Specifically, we employ a site-based SGMM defined in the frequency domain \cite{Broccardo2017, Vlachos2016}. The model is the spectral representation of the original time-domain model implemented in \cite{Rezaeian2010}. It targets broad-band excitations, which are typically associated with far-field ground motions. 

In detail, the SGMM is completely characterized by an evolutionary power spectral density (EPSD) \cite{Priestley1965}. Like its original time-domain counterpart, this representation allows separating the temporal and spectral components of the process \cite{Broccardo2017, Vlachos2016}. In this study, without losing generality, we neglect the non-stationary spectral characteristics of the ground motion. In fact, within a good engineering approximation, the frequency content and the bandwidth of the strong ground motion phase can be assumed constant for broad-band excitations. Moreover, it is assumed that severe structural damage occurs during the strong motion phase.

Finally, the spectral content of the process is represented by a normalized stationary Kanai--Tajimi power spectral density (KT-PSD), which is a function of two parameters: the main frequency, $\omega_g$, and the bandwidth, $\zeta_g$. The normalized KT-PSD produces a stationary process with unit variance so that the intensity of the ground motion is completely controlled by a time-modulating function. This model is a common choice in earthquake engineering to describe broad-band far-field ground motions. We use a gamma modulating function (\cite{Rezaeian2010,Broccardo2017}), which is completely defined by the expected Arias intensity $I_a$, the time at which 45\% of the expected Arias intensity is reached, $t_{\mathrm{mid}}$, and the effective duration of the motion, $D_{\mathrm{5-95}}$. Finally, the complete SGMM EPSD is given by modulating the normalized KT-PSD with the time-modulating function. Moreover, to ensure zero residual velocity and displacement, we apply a high-pass filter using the evolutionary theory of Priestley (see \cite{Broccardo2017} for a detailed description).  To summarize, the SGMM model parameters are $\ve{x} = [I_a, t_{\mathrm{mid}}, D_{\mathrm{5-95}},  \omega_{g}, \zeta_g]$.

Next, we fit the SGMM model to a catalog of recorded far-field ground motions from the PEER NGA-West2 database (the same used in \cite{broccardo_dab2019}). The catalog includes 71 ground motions recorded at a range of distances (10-90 km) and site conditions from reverse earthquakes with a magnitude between 6 and 7.6. The two horizontal components of each record are rotated into the major and intermediate principal directions (\cite{Rezaeian2010}). In this study, we used only the major component (i.e., we used 71 time series).
The fitting procedure for the frequency content of the ground motion is described in detail in Broccardo and Dabaghi \cite{Broccardo2017}. However, in Broccardo and Dabaghi \cite{Broccardo2017}, $\omega_g$ and $\zeta_g$ are a time-varying function, while in this study $\omega_g$ corresponds to the main frequency of the ground motions at $t_{\mathrm{mid}}$ (which is considered the strong phase of the ground motion). Moreover, we fix $\zeta_g$ to a constant value of 0.9, which was a good approximation for the selected broad-band excitations\footnote{We found that the EDP response was not sensitive to large values of $\zeta_g$. Therefore, we used a point approximation and reduced the parameter space. Note that this approximation does not limit the generality of the SPCE approach for fragility model computation.}. The approach to estimates the parameters $I_a, t_{\mathrm{mid}}, D_{\mathrm{5-95}}$ follows \cite{Rezaeian2010}. In this respect, the free SGMM parameters are random variables (i.e., $\ve{x}$ becomes a random vector, $\ve{X}$) to account for the epistemic uncertainty related to the chosen data set.

Provided with the 71 estimates of the free parameters, we  fit a joint-probability model based on log-normal marginal distributions and a Gaussian copula (i.e., a joint log-normal distribution). Consequently, the models also account for the dependence structure among the parameters. The joint-probability model parameters are reported in Table \ref{tab:gm_joint_prob}. Finally, the simulation of the synthetic time series follows a two-step simulation (which is typical in a stochastic simulator setting). First, the SGMM parameters are sampled from the joint log-normal distribution. Second, using the synthesis formula of the frequency domain representation of a stochastic process \cite{Shinozuka1991}, the time series are generated by filtering white noise Gaussian vectors with the EPSD and the high-pass filter. Therefore, for a given set of model parameters $\ve{X}=\ve{x}$, multiple time series can be generated. Consequently, the EDP of interest is a random variable even when $\ve{X}=\ve{x}$.

\begin{table}[!ht]
	\centering
	\caption{Ground motion parameters, g is the gravitational constant expressed in [m/s$^2$]}
	\begin{tabular}{lc}
		\toprule
		Name    & Distribution \\
		\midrule
		$I_a$~[g$^2\cdot$ s] & $\cl\cn\left(-4.61,1.45^2\right)$ \\
		$t_{\mathrm{mid}}$~[s] &  $\cl\cn\left(2.55,0.90^2\right)$    \\
		$D_{\mathrm{5-95}}$~[s]  &  $\cl\cn\left(2.67,0.53^2\right)$   \\
		$\omega_g$~[rad/s] & $\cl\cn\left(1.42,0.59^2\right)$ \\
		\midrule
		\thead{Correlation \\ matrix} & \thead{$R = \begin{pmatrix}
				1& 0.015 & -0.23 & -0.13 \\
				0.015 & 1& 0.68 & -0.36 \\
				-0.23 & 0.68 & 1 & -0.11 \\
				-0.13 & -0.36 & -0.11& 1\end{pmatrix}$}   \\
		\bottomrule
	\end{tabular}
	\label{tab:gm_joint_prob}
\end{table}

\subsection{Toy example}
\label{sec:ex1}
\noindent In this example, we introduce the properties of a three-story shear frame idealized as a three-degree of freedom system. We are interested in the dynamic response of the system subjected to the ground motions generated according to \Cref{sec:sgm}. The interstory behavior is inelastic, with a force-interstory-drift relationship based on a Bouc-Wen hysteretic model \cite{wen1976method}. Specifically, the $i$-th interstory restoring force is written as
\begin{equation}
	q_i(v_i(t),\dot v_i(t)) = k_i\left[\alpha v_i(t) + (1-\alpha)z(t)\right],
\end{equation}
where $v_i(t)$ denotes the interstory drift, $\alpha$ is a parameter that controls the degree of inelasticity (i.e., $\alpha =1$ corresponds to the linear case), $k_i$ is the initial elastic interstory stiffness, and $z(t)$ is the hysteretic response governed by the following law
\begin{equation}
	\dot z(t) = -\gamma\left|\dot v(t)\right|\left|z(t)\right|^{n-1}-\eta\left|z(t)\right|^n\dot v_i(t)+A\dot v(t),
\end{equation}
where $\gamma, n, A$ and $\eta$ are the model parameters. The values of structural properties, including the local masses $m_i$ and damping $c_i$, and model parameters are reported in Table \ref{tab:sp_BWp}. The story yield displacement, $\delta_y$, is set to 0.01~m and the post-hardening stiffness is set at 10\% of the elastic stiffness $k_i$ for all the three stories. The EDP of interest is the maximum interstory drift, i.e., 
\begin{equation}
	Y = \max\left[\max_t[v_1(t)],\max_t[ v_2(t)],\max_t[v_3(t)]\right].
\end{equation}

\begin{table*}[!htbp]
	\centering
	\caption{Structural properties and Bouc-Wen parameters ($\delta_y = 0.01$~m).}
	\begin{tabular}{cccccccccc}
		\toprule
		& $m_i$ $10^6$[kg] & $c_i$ [$10^6$[Ns/m] & $k_i$ $10^8$[N/m] & $\alpha$ &  $n$ & $\gamma$ [1/m$^n$] & $\eta$ [1/m$^n$] & $A$ \\
		\midrule
		Story 1 & 1 & 1.73 & 3.0 & 0.1 &   5 & $1/(2\delta_y)^n$& $1/(2\delta_y)^n$ & 1 \\
		Story 2 & 1 & 1.73 & 2.4 & 0.1 &   5 & $1/(2\delta_y)^n$& $1/(2\delta_y)^n$ & 1 \\
		Story 3 & 1 & 1.73 & 1.5 & 0.1 &   5 & $1/(2\delta_y)^n$& $1/(2\delta_y)^n$ & 1\\
		\bottomrule
	\end{tabular}
	\label{tab:sp_BWp}
\end{table*}

\begin{figure}[!htbp]
	\centering
	\begin{subfigure}{.48\linewidth}
		\centering
		\includegraphics[height=0.6\linewidth, keepaspectratio]{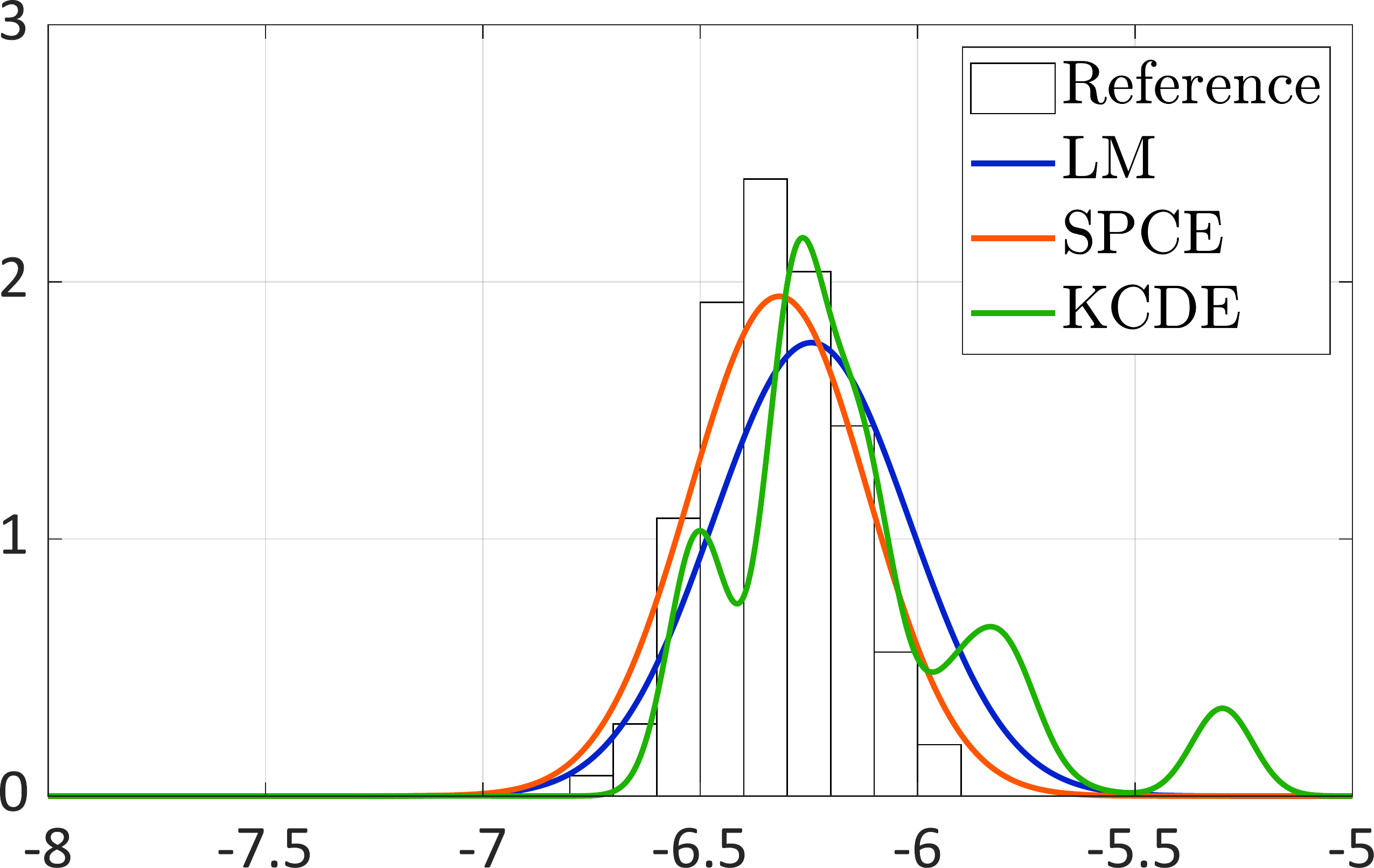}
		\caption{$\ve{x} = (0.0013,69.94,61.59,5.66)$}
		\label{fig:Ex1GMPpdf1}
	\end{subfigure}
	\hfill
	\begin{subfigure}{.48\linewidth}
		\centering
		\includegraphics[height=0.6\linewidth, keepaspectratio]{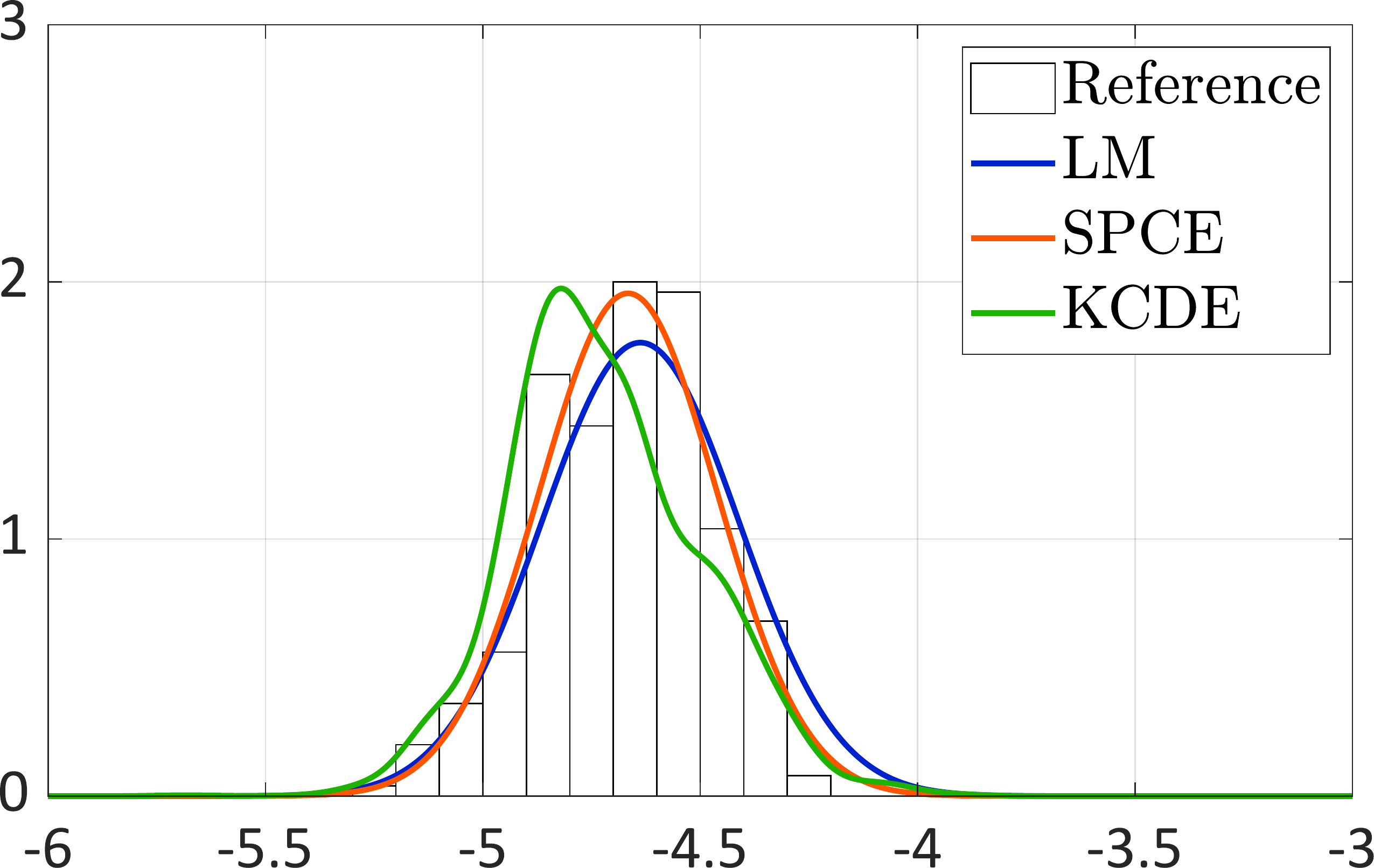}
		\caption{$\ve{x} = (0.017,7.57,10.06,7.01)$}
		\label{fig:Ex1GMPpdf2}
	\end{subfigure}
	\begin{subfigure}{.48\linewidth}
		\centering
		\includegraphics[height=0.6\linewidth, keepaspectratio]{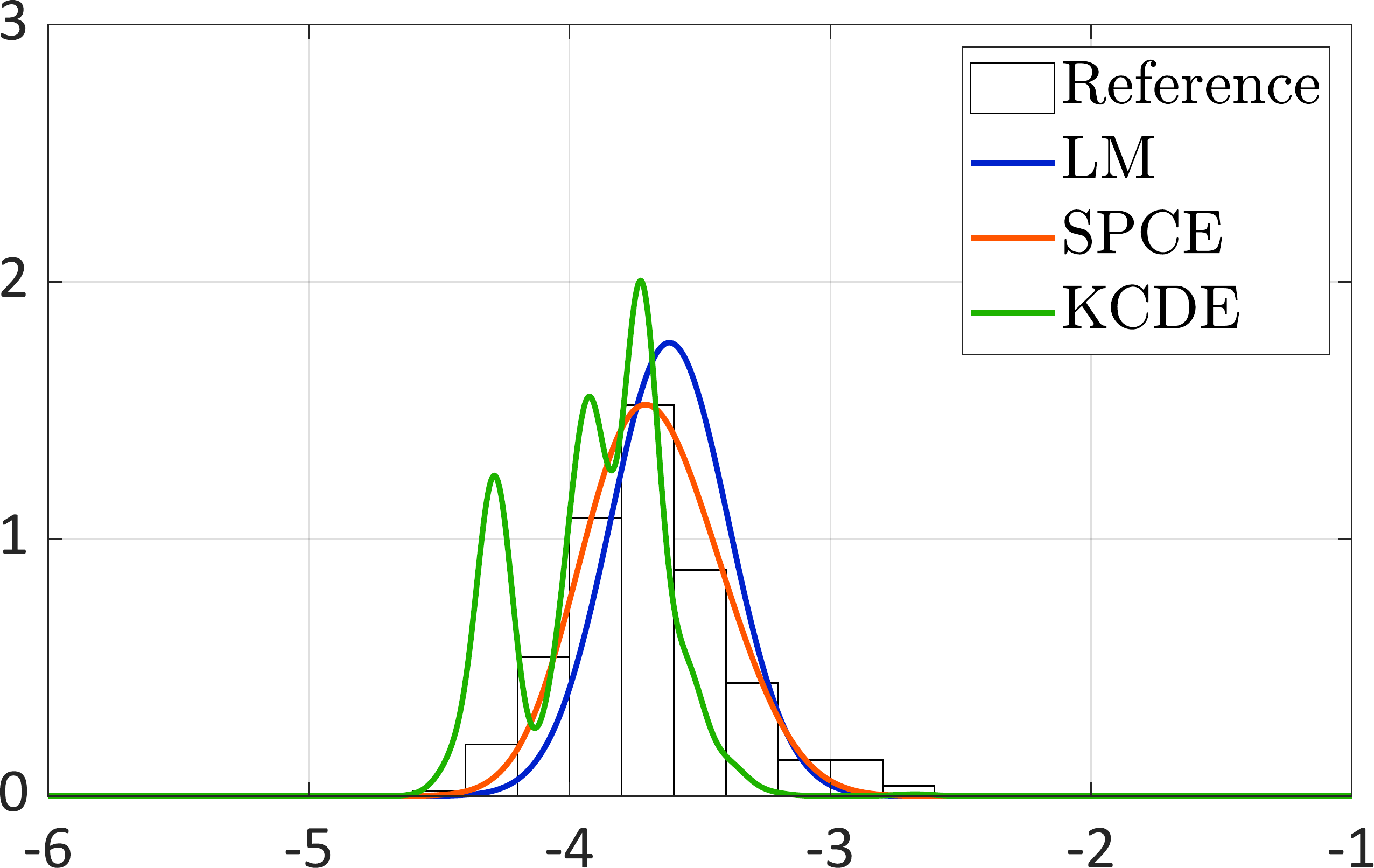}
		\caption{$\ve{x} = (0.055,3.67,3.45,4.79)$}
		\label{fig:Ex1GMPpdf3}
	\end{subfigure}
	\hfill
	\begin{subfigure}{.48\linewidth}
		\centering
		\includegraphics[height=0.6\linewidth, keepaspectratio]{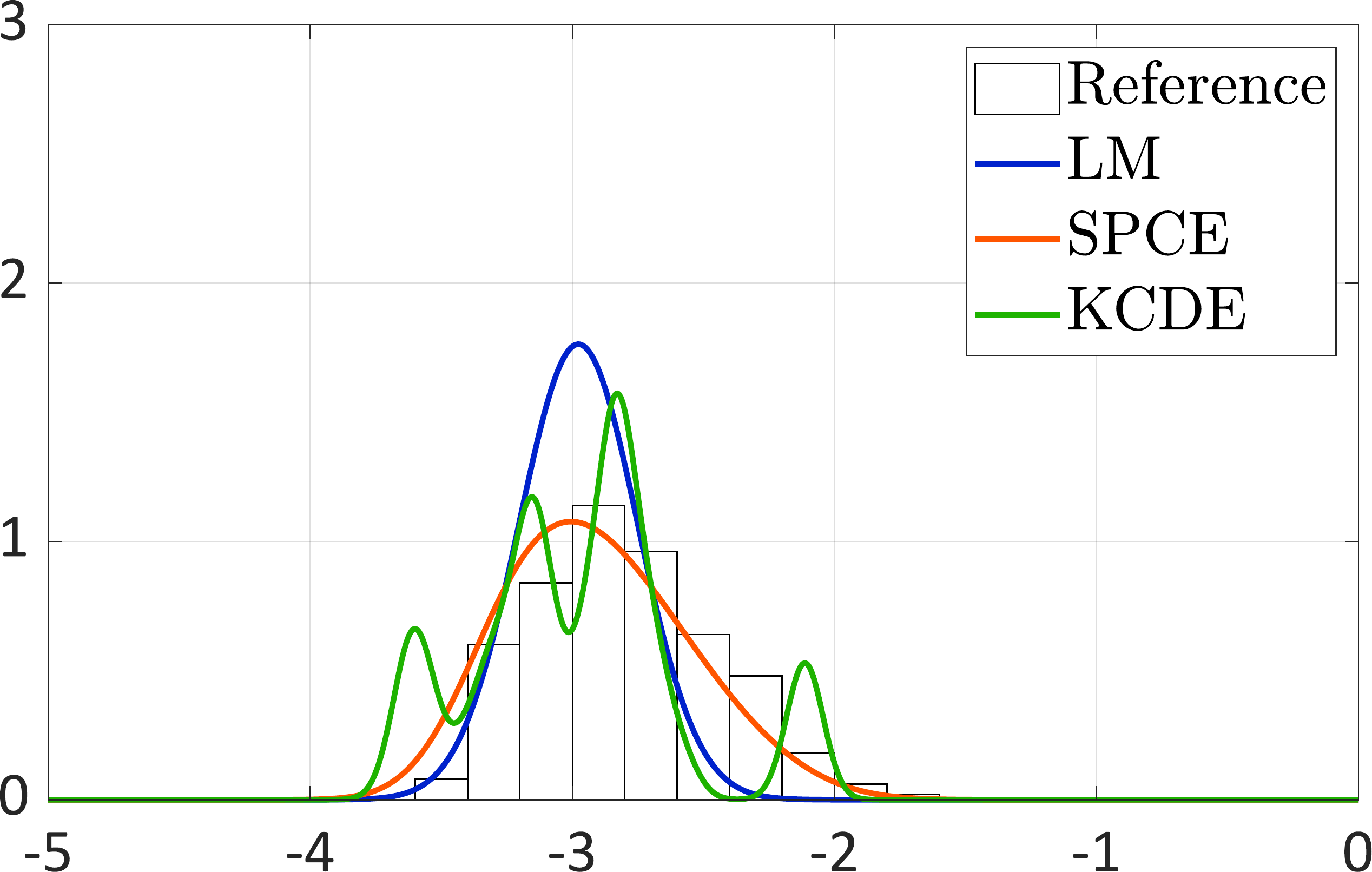}
		\caption{$\ve{x} = (0.16,7.23,6.15,2.70)$}
		\label{fig:Ex1GMPpdf4}
	\end{subfigure}
	\caption{Example 1 --- comparison of emulated PDFs of $\log\left(Y_{\ve{x}}\right)$ for four different values of $\ve{x}$; the models are built on $N=1{,}000$ simulations.}
	\label{fig:Ex1GMPpdf}
\end{figure}

\Cref{fig:Ex1GMPpdf} illustrates the conditional PDF of the maximum interstory drift for four different values of the ground motion variables. The models are constructed based on a total number of $1{,}000$ simulations. The reference histograms are obtained by replicating the simulation $250$ times for each set of ground motion parameters, i.e., we generated $250$ ground motions for each $\ve{x}$ and computed the associated structural responses. The distributions are plotted on the logarithmic transform of $Y_{\ve{x}}$, which allows for verifying the assumptions of the linear model. 

As shown in \Cref{fig:Ex1GMPpdf}, the linear model can represent the overall location and shape of the conditional distribution: the prediction of the mean values are close to the reference histograms that demonstrate normal-like shapes. Nevertheless, the linear model loses some details of the mean estimation in \Cref{fig:Ex1GMPpdf1,fig:Ex1GMPpdf3} and cannot capture the heteroskedastic effect (varying variance). The PDF predictions of KCDE are quite poor, as it yields spurious oscillations in \Cref{fig:Ex1GMPpdf1,fig:Ex1GMPpdf3,fig:Ex1GMPpdf4}. This is because the bandwidth selection procedure \cite{Hayfield2008} is designed for estimating the conditional CDF. Moreover, the conditional distribution estimation requires estimating the joint distribution of $(\ve{X},Y)$ in \cref{eq:nonparam} which is of dimension $5$. This is rather high for nonparametric estimators and leads to the observed poor predictions. In contrast, SPCE can accurately emulate the PDFs in terms of not only the location and the heteroskedastic effect but also the shape of the distributions: \Cref{fig:Ex1GMPpdf4} is slightly right-skewed which is well represented by SPCE.

To study the convergence of the various methods, we generated a big data pool of size $10^5$ (following the distribution of $\ve{X}$ described in \Cref{tab:gm_joint_prob}). We randomly subsampled it to have samples of desired sizes $N \in \acc{250;500;1{,}000;2{,}000;4{,}000}$ to train the models. Note that this mimics the procedure of random design of experiment. To account for the uncertainties in the estimation, we repeated the procedure 20 times for each sample size (i.e., we obtain 20 models constructed on independent subsamples for each $N$). To evaluate the error metrics defined in \cref{eq:Rlevel1,eq:errorq}, we generated a validation set of size $400$. For each validation point, we used $250$ replications to have a reference distribution (meaning a total number of $400\times250$ simulations for the validation set). The error estimates for each sample size are represented by box plots constructed from the 20 repetitions of the full analysis. 

\begin{figure}[htbp!]
	\centering
	\includegraphics[width=0.75\linewidth]{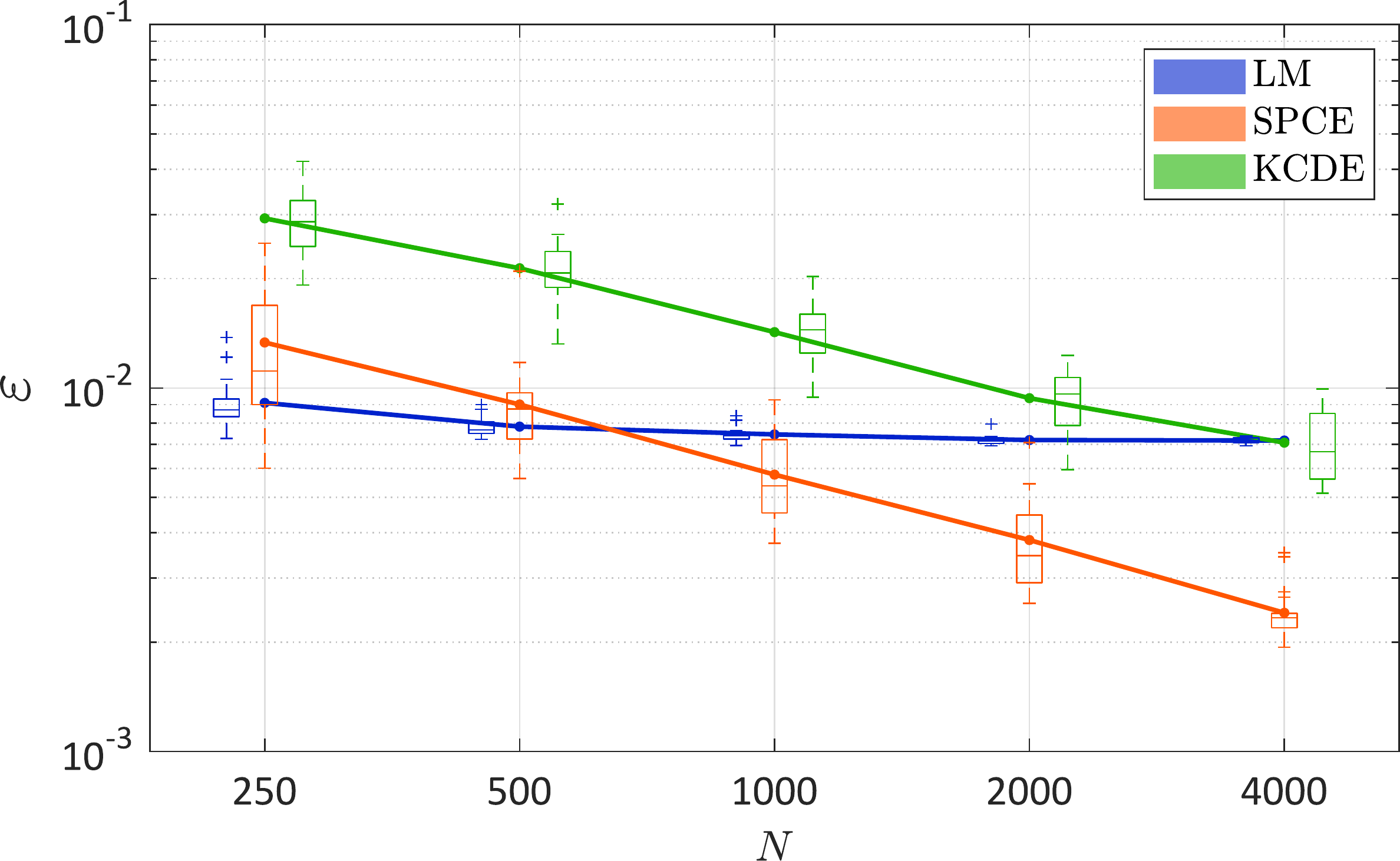}
	\caption{Example 1 --- comparison of the convergence among the models in terms of the normalized Wasserstein distance. The lines correspond to the average values over 20 repetitions of the full analysis, whereas the box plot summarizes the 20 results.}
	\label{fig:Ex1WS_GMP}
\end{figure}

\Cref{fig:Ex1WS_GMP} shows the results of the models for estimating the conditional distribution. For relatively small sample sizes $N\leq 500$, the linear model gives the best results. This is because the linear model is very simple, and its assumptions are relatively ``suitable'' for this example. More precisely, the error of a statistical model can be decomposed into bias and variance \cite{James2014}. In \Cref{fig:Ex1GMPpdf}, we observe that the conditional distribution is close to Gaussian, the mean function does not exhibit a strong nonlinearity, and the heteroskedastic effect is relatively weak. Therefore, the bias of the linear model is rather small. Because of its simplicity, the linear model has a small variance. As a result, when only a few data points are available, the linear model gives the best results. However, with increasing sample size, the errors of the linear model run into a plateau. This is due to the irreducible bias (caused by the model misspecification). On the contrary, SPCE and KCDE are more flexible models that have smaller biases but bigger variances. Hence, both models exhibit a clear decay of the error. Due to its nonparametric feature, KCDE is merely comparable to the linear model for $N = 4{,}000$. When enough samples are available, i.e., $N\geq 1{,}000$, SPCE is the best model. Furthermore, the average error of SPCE is three times smaller than those of the linear model and kernel estimator for $N= 4{,}000$.

\begin{figure}[htbp!]
	\centering
	\begin{subfigure}{.48\linewidth}
		\centering
		\includegraphics[height=0.78\linewidth, keepaspectratio]{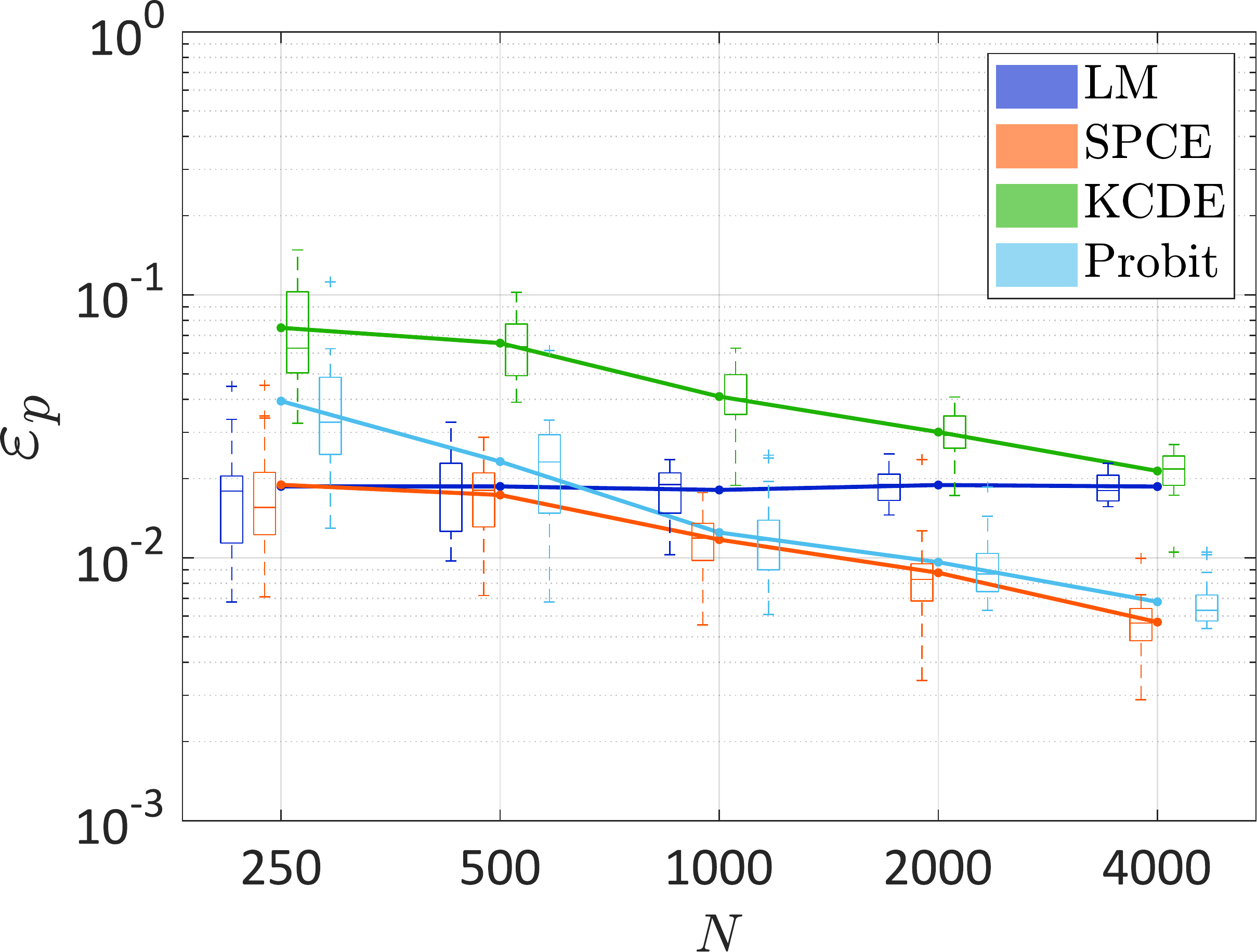}
		\caption{Fragility function for $\delta_0=0.02$~m}
		\label{fig:Ex1GMPFA1}
	\end{subfigure}
	\hfill
	\begin{subfigure}{.48\linewidth}
		\centering
		\includegraphics[height=0.78\linewidth, keepaspectratio]{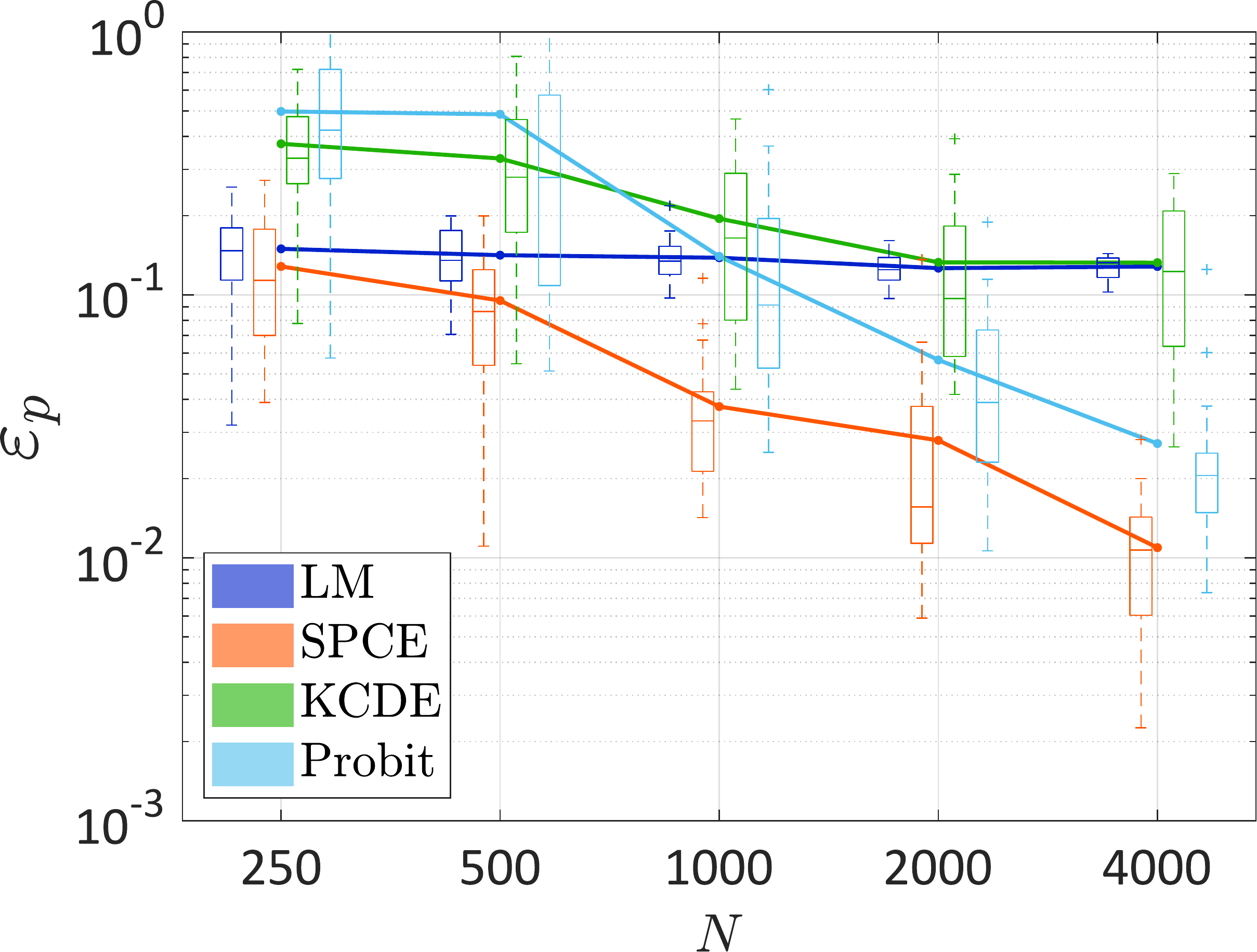}
		\caption{Fragility function for $\delta_0=0.07$~m}		\label{fig:Ex1GMPFA2}
	\end{subfigure}
	\caption{Example 1 --- comparison of the convergence among the models in terms of the fragility functions. The lines correspond to the average values over 20 repetitions of the full analysis, whereas the box plot summarizes the 20 results.}
	\label{fig:Ex1FA_GMP}
\end{figure}

When considering fragility functions, we select two thresholds $\delta_0 = 0.02$~m and $\delta_0 = 0.07$~m. The relative mean-squared errors for estimating the associated fragility functions are reported in \Cref{fig:Ex1FA_GMP}. In general, SPCE produces the best overall approximation of the fragility functions for all sample sizes. Similar to what we observed in \Cref{fig:Ex1WS_GMP}, the performance of the linear model barely improves with increasing $N$. For $\delta_0 = 0.07$~m, SPCE outperforms the linear model in the case of a few samples $N\leq 500$. This indicates that SPCE better approximates the tails. The probit model yields relatively large errors for training sets of sizes $N\leq1{,}000$ in the estimation of the fragility function associated with $\delta_0 = 0.07$~m. This is because this model ignores the precise values of EDP and only works with the binary variable. For $\delta_0 = 0.07$~m, only a small fraction of samples (about $1.3\%$) exceed the threshold. Consequently, the probit model only produces reliable estimates for large $N$. Finally, KCDE performs quite poorly even though the associated bandwidth selection procedure is designed for CDF estimation.

\subsection{Three-story frame}
\label{sec:ex2}
\begin{figure}[!htbp]
	\centering
	\begin{subfigure}[t]{.48\linewidth}
		\centering
		\includegraphics[height=0.7\linewidth, keepaspectratio]{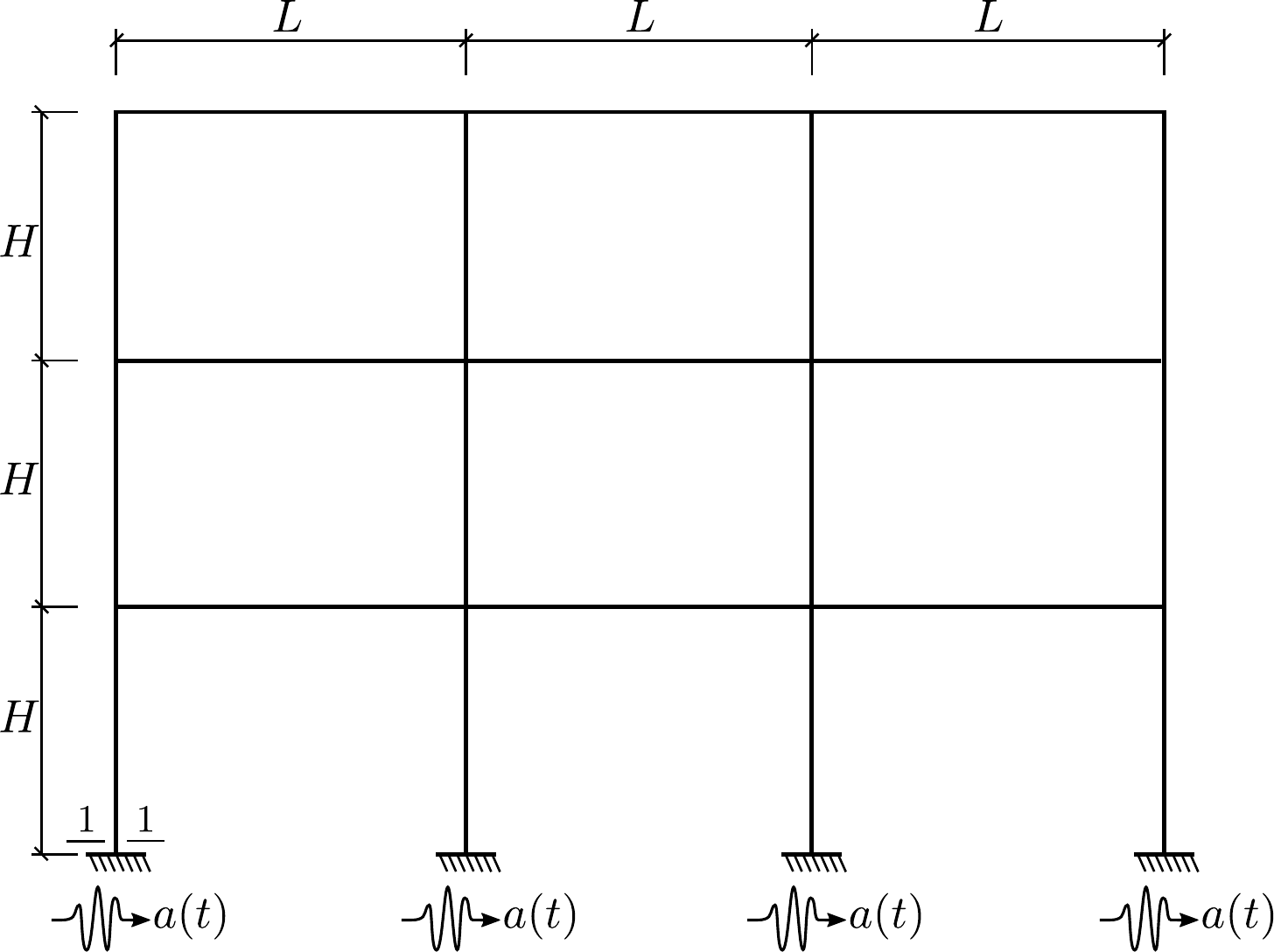}
		\caption{Illustration of the frame structure}
		\label{fig:frame}
	\end{subfigure}
	\hfill
	\begin{subfigure}[t]{.48\linewidth}
		\centering
		\includegraphics[height=0.7\linewidth, keepaspectratio]{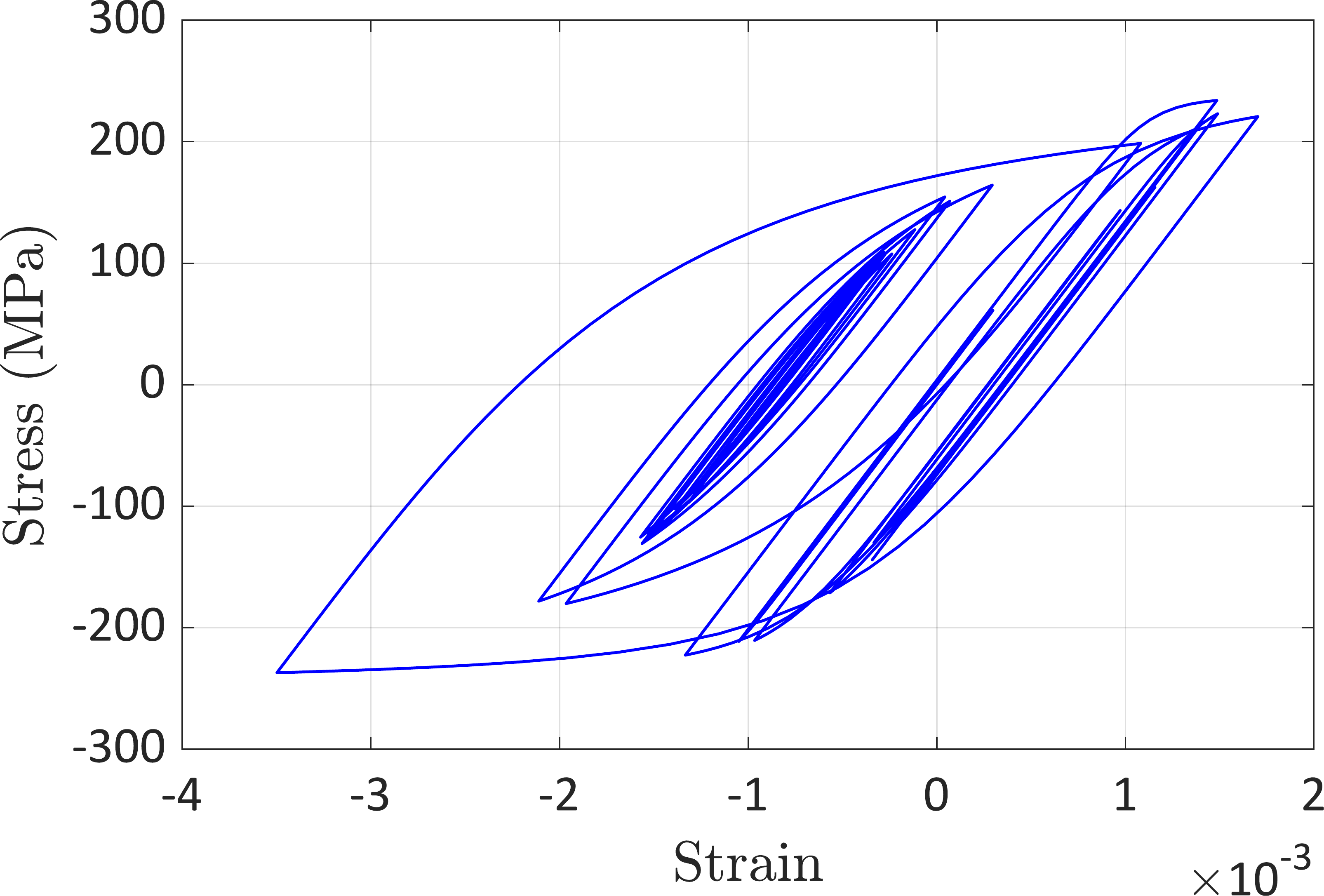}
		\caption{Hysteric behavior of the steel material at section 1-1 for a ground motion}
		\label{fig:stress-strain}
	\end{subfigure}
	\caption{Example 2 --- three-story steel frame.}
	\label{fig:Ex2illust}
\end{figure}

\noindent 
As a second example, we apply the methods to study the three-story steel frame modeled with the software OpenSees \cite{OpenSees}. The geometry of the structure is shown in \Cref{fig:frame}, and the story height and floor span are $H = 3$~m and $L = 5$~m, respectively. We choose the standard European IPE~A~330 for the beams and HE~200~AA for the columns. 

The mechanical property of the steel follows the uniaxial Giuffre-Menegotto-Pinto model with isotropic strain hardening (material of type ``Steel02'' in OpenSees). More precisely, we set the Young's modulus to $E = 205{,}000$~MPa, the yield stress to $f_y = 235$~MPa, and the strain hardening ratio to $b = 0.01$ (the other parameters controlling the elastic-plastic transition are given by $R0 = 18$, $CR1 = 0.925$, and $CR2 = 0.15$). The load applied to the structure consists of dead load (weight of frame elements and supported floors) and live load, which results in a total distributed load on the beams equal to $q = 20$~kN/m \cite{Mai2017}. 

The structural components (beams and columns) are modeled by nonlinear beam elements based on the iterative force-based formulation. The element cross-sections are defined by a set of fiber sections, which allows modeling the plasticity over the cross-section. \Cref{fig:stress-strain} illustrates the stress-strain relation of the bottom left column for the frame under an example ground motion. The first two fundamental periods of the structure are 0.950~s and 0.317~s (from modal analysis), respectively. In this study, we are interested in the dynamic response of the system subjected to the ground motions generated according to \Cref{sec:sgm}. The EDP of interest is the maximum interstory drift ratio.

\begin{figure}[!htbp]
	\centering
	\begin{subfigure}{.48\linewidth}
		\centering
		\includegraphics[height=0.6\linewidth, keepaspectratio]{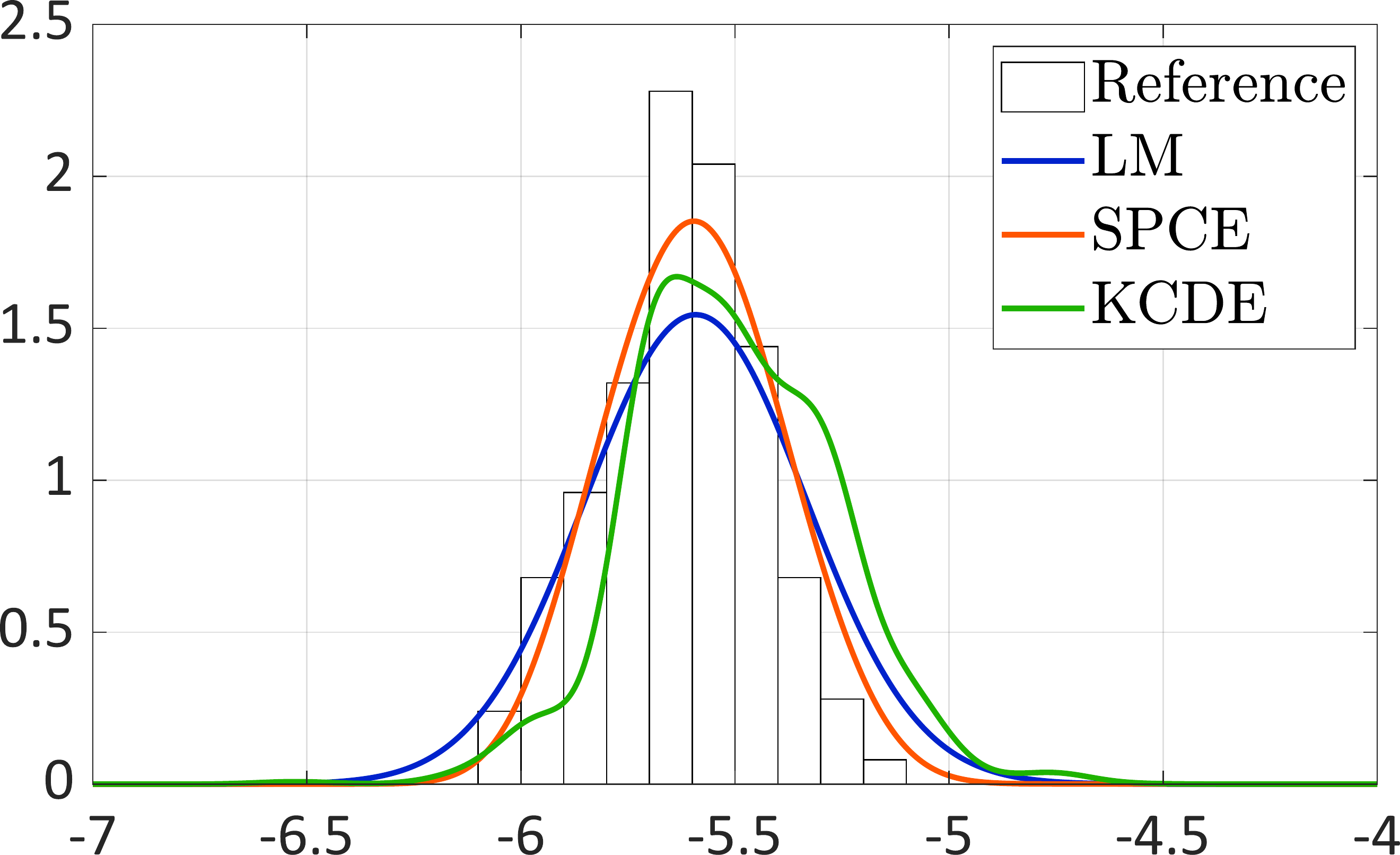}
		\caption{$\ve{x} = (0.0098,44.21,47.08,2.33)$}
		\label{fig:Ex2GMPpdf1}
	\end{subfigure}
	\hfill
	\begin{subfigure}{.48\linewidth}
		\centering
		\includegraphics[height=0.6\linewidth, keepaspectratio]{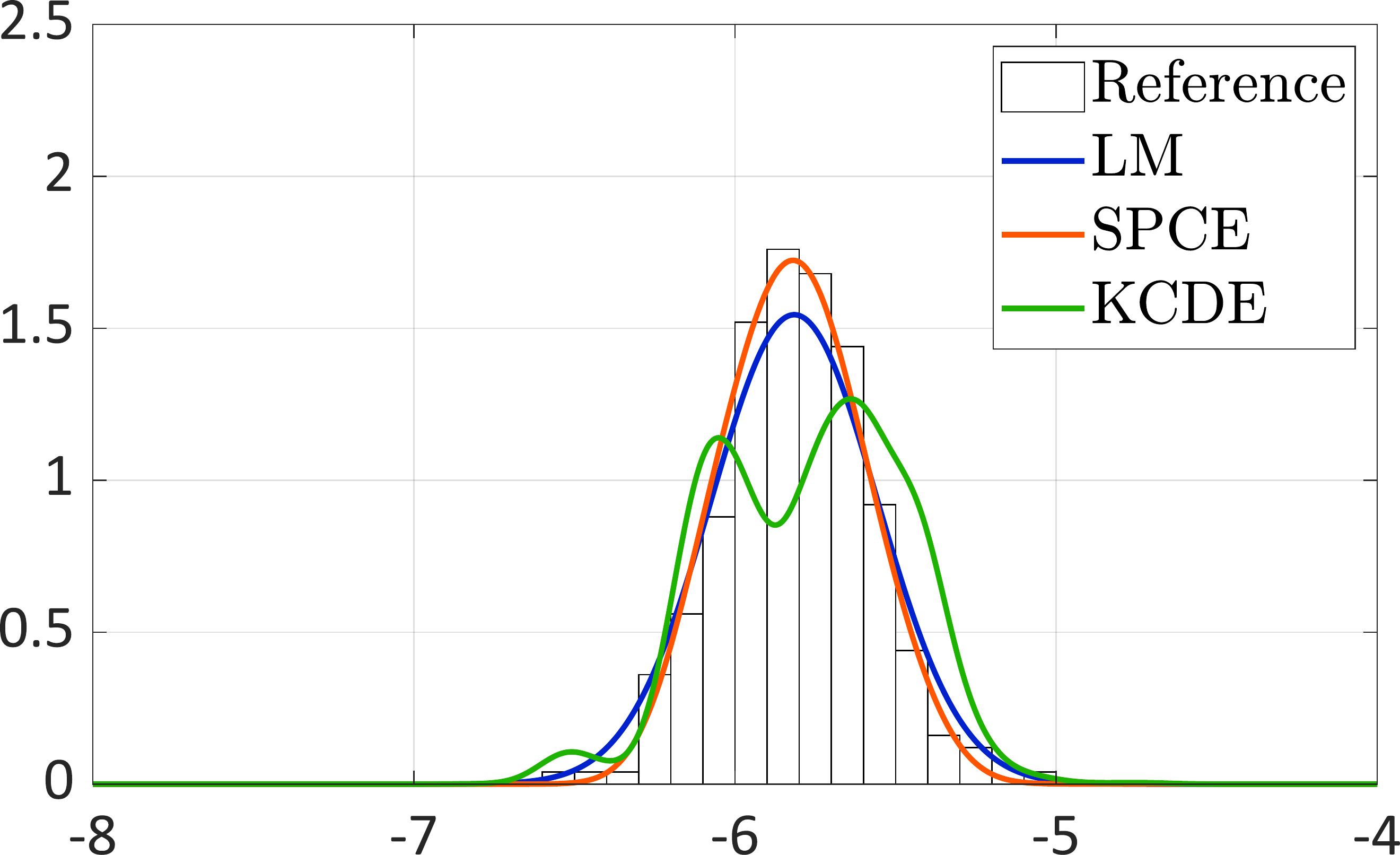}
		\caption{$\ve{x} = (0.0058,31.36,26.89,2.69)$}
		\label{fig:Ex2GMPpdf2}
	\end{subfigure}
	\begin{subfigure}{.48\linewidth}
		\centering
		\includegraphics[height=0.6\linewidth, keepaspectratio]{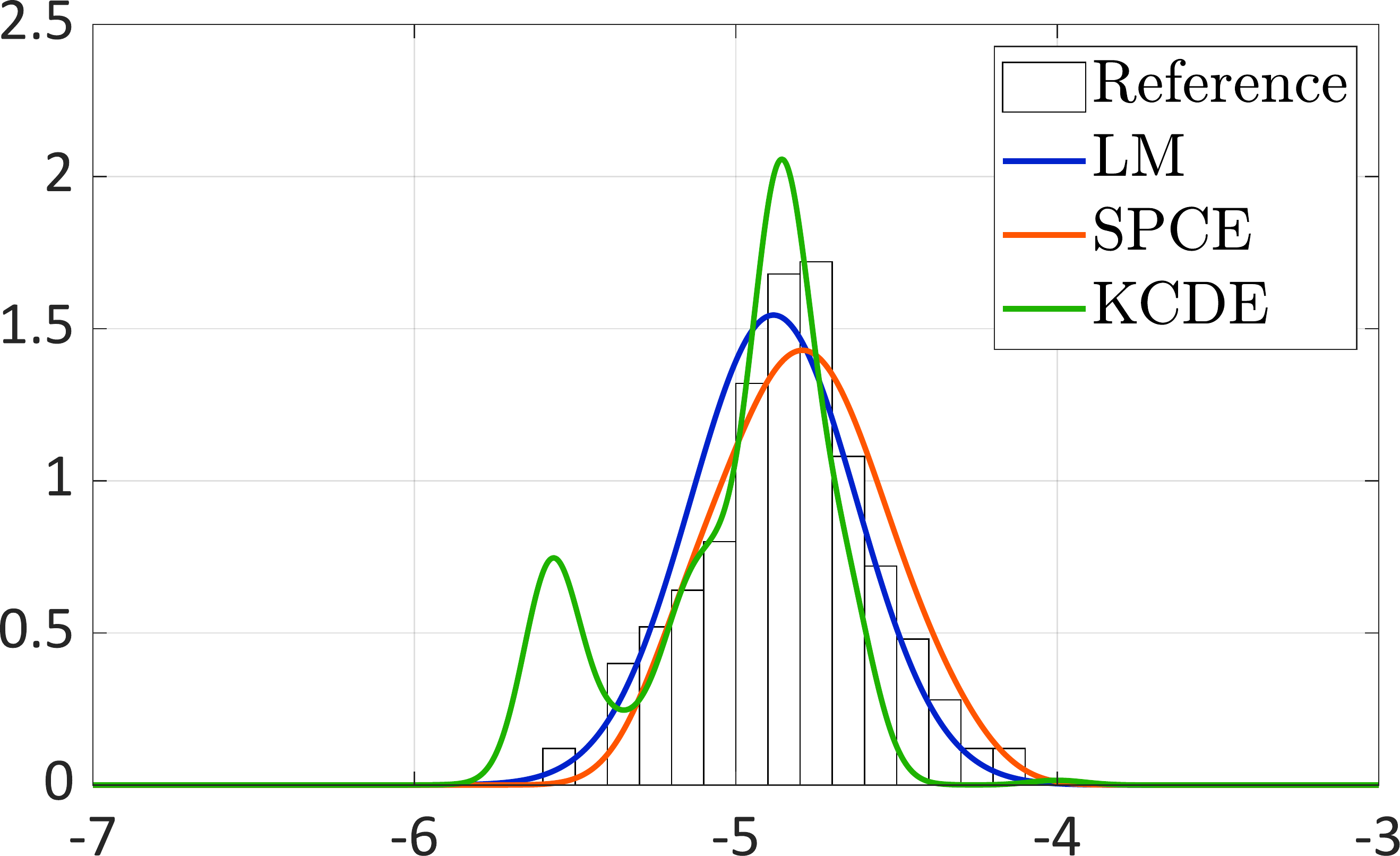}
		\caption{$\ve{x} = (0.014,15.15,10.19,1.24)$}
		\label{fig:Ex2GMPpdf3}
	\end{subfigure}
	\hfill
	\begin{subfigure}{.48\linewidth}
		\centering
		\includegraphics[height=0.6\linewidth, keepaspectratio]{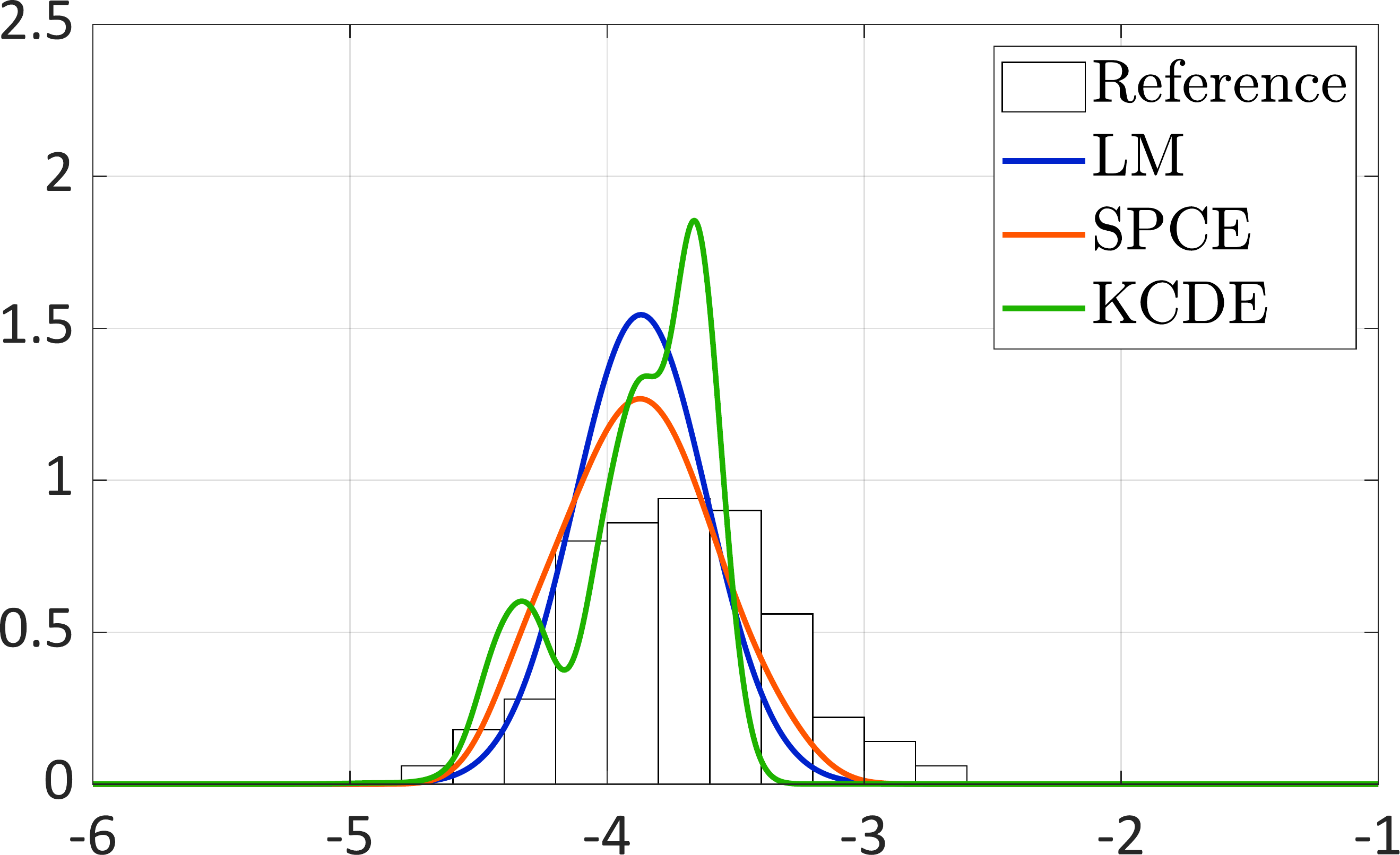}
		\caption{$\ve{x} = (0.11,12.85,5.49,1.76)$}
		\label{fig:Ex2GMPpdf4}
	\end{subfigure}
	\caption{Example 2 --- comparison of emulated PDFs of $\log\left(Y_{\ve{x}}\right)$ for four different values of $\ve{x}$; the models are built on $N=1{,}000$ simulations.}
	\label{fig:Ex2GMPpdf}
\end{figure}

\Cref{fig:Ex2GMPpdf} shows the prediction of the conditional PDFs for four different values of $\ve{x}$. The reference histogram of each $\ve{x}$ is calculated by performing $250$ replications, and the surrogate models are built on $1{,}000$ simulations. Similar to the first example \Cref{fig:Ex1GMPpdf}, we observe that the conditional distributions have bell shapes that are close to Gaussian distributions. The linear model can well approximate the location of the distributions, so the (log-)mean function does not demonstrate a strong non-linearity. The variance of the conditional distribution does not vary too much. The linear model shows a good overall approximation, but it fails to characterize the precise variation of the distribution. On the contrary, the kernel method is too flexible and completely mispredicts the shape of the distribution. In contrast, SPCE turns out to accurately represent not only the location and shape of the distribution but also the heteroskedastic effect.

For the convergence study, we followed the same procedure as \Cref{sec:ex1}. In this example, we generated a data pool of size $50{,}000$. We randomly subsampled this data set to have the experimental design of sizes $N\acc{250;500;1{,}000;2{,}000;4{,}000}$ to build the surrogate models. We repeated the analysis 20 times for each value of $N$ to account for the uncertainties (due to the random ground motion parameters and the intrinsic stochasticity of the ground motion model). To evaluate the error defined in \cref{eq:Rlevel1,eq:errorq}, we created a validation set of size $200$, and we performed $250$ replications for each validation point to have a reference conditional distribution.

\begin{figure}[htbp!]
	\centering
	\includegraphics[width=0.75\linewidth]{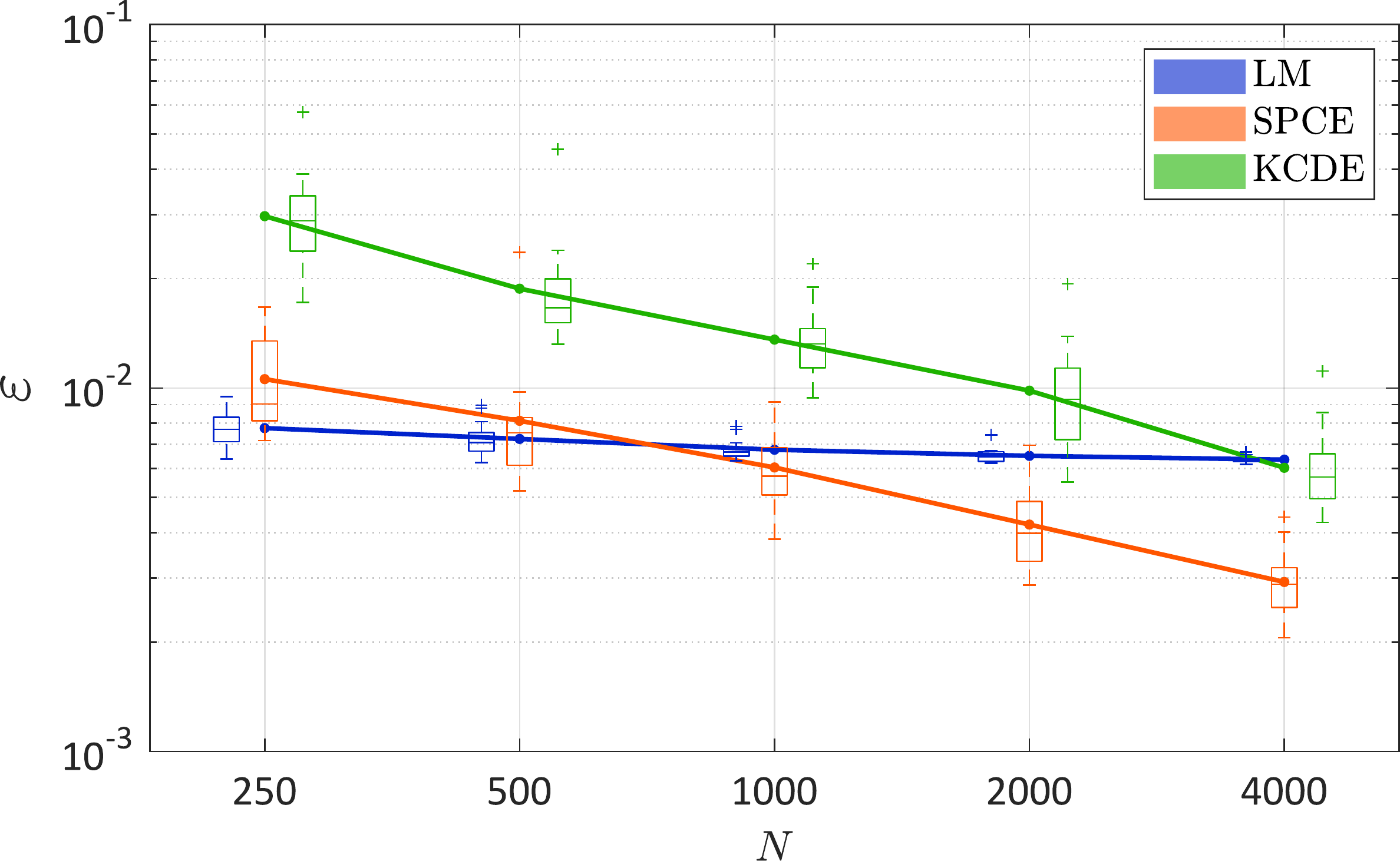}
	\caption{Example 2 --- comparison of the convergence among the models in terms of the normalized Wasserstein distance. The lines correspond to the average values over 20 repetitions of the full analysis, whereas the box plot summarizes the 20 results.}
	\label{fig:Ex2WS_GMP}
\end{figure}

\Cref{fig:Ex2WS_GMP} shows the error metric defined in \cref{eq:Rlevel1}. Similar to \Cref{fig:Ex1WS_GMP}, the linear model is superior to SPCE and KCDE when only $N = 250$ data points are used. With increasing $N$, its errors exhibit narrower spreads, but the average values do not decrease due to the bias resulting from the model simplicity. The kernel estimator exhibits a better convergence rate but performs poorly overall. SPCE has a similar performance to the linear model at $N = 500$ and surpasses the latter for $N\geq 1{,}000$. In addition, SPCE has a clear decay of the errors with a similar rate to KCDE. For $N = 4{,}000$, the average error of SPCE is less than half of those of the linear model and KCDE.

\begin{figure}[htbp!]
	\centering
	\begin{subfigure}{.48\linewidth}
		\centering
		\includegraphics[height=0.75\linewidth, keepaspectratio]{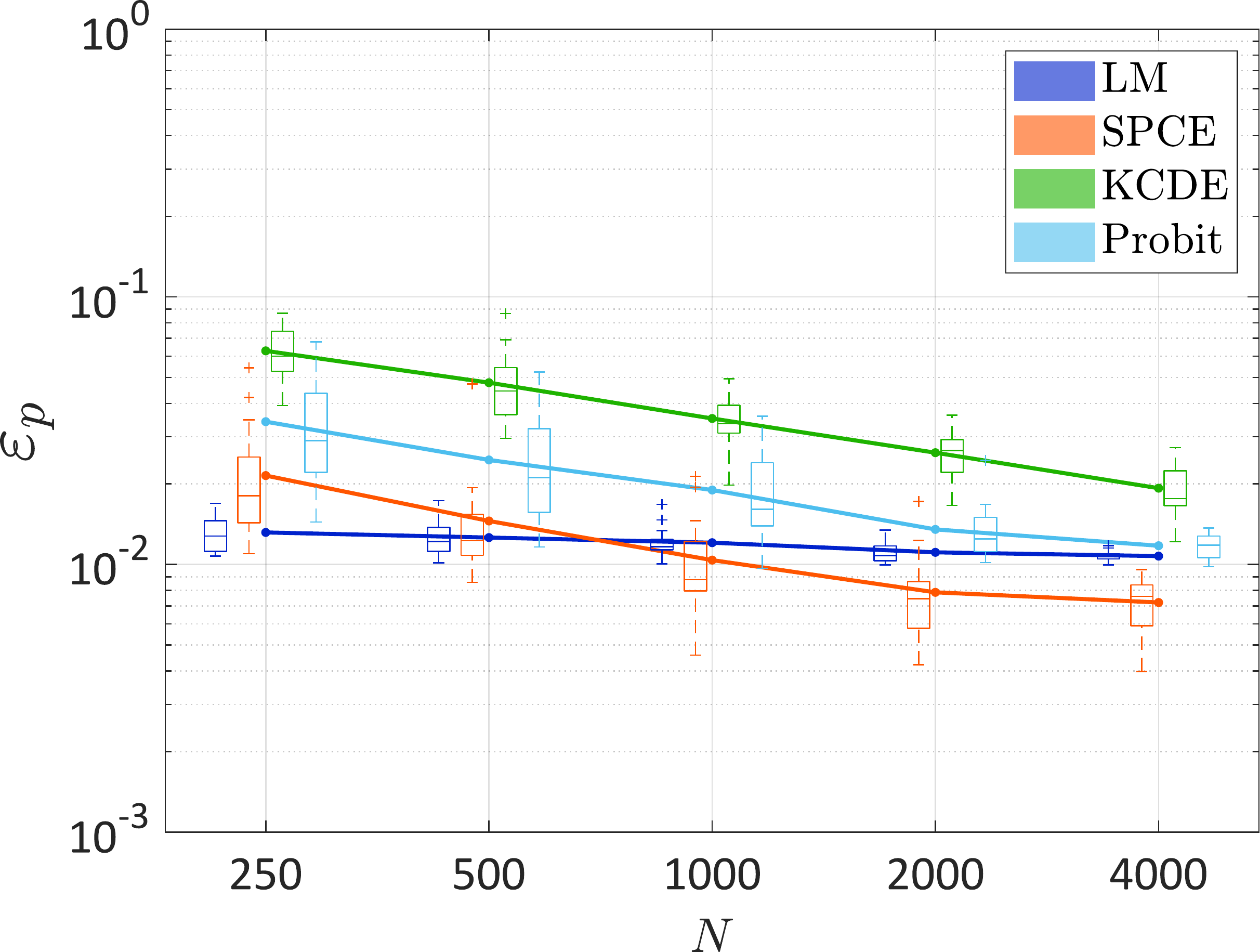}
		\caption{Fragility function for $\delta_0=0.7\%$}
		\label{fig:Ex2GMPFA1}
	\end{subfigure}
	\hfill
	\begin{subfigure}{.48\linewidth}
		\centering
		\includegraphics[height=0.75\linewidth, keepaspectratio]{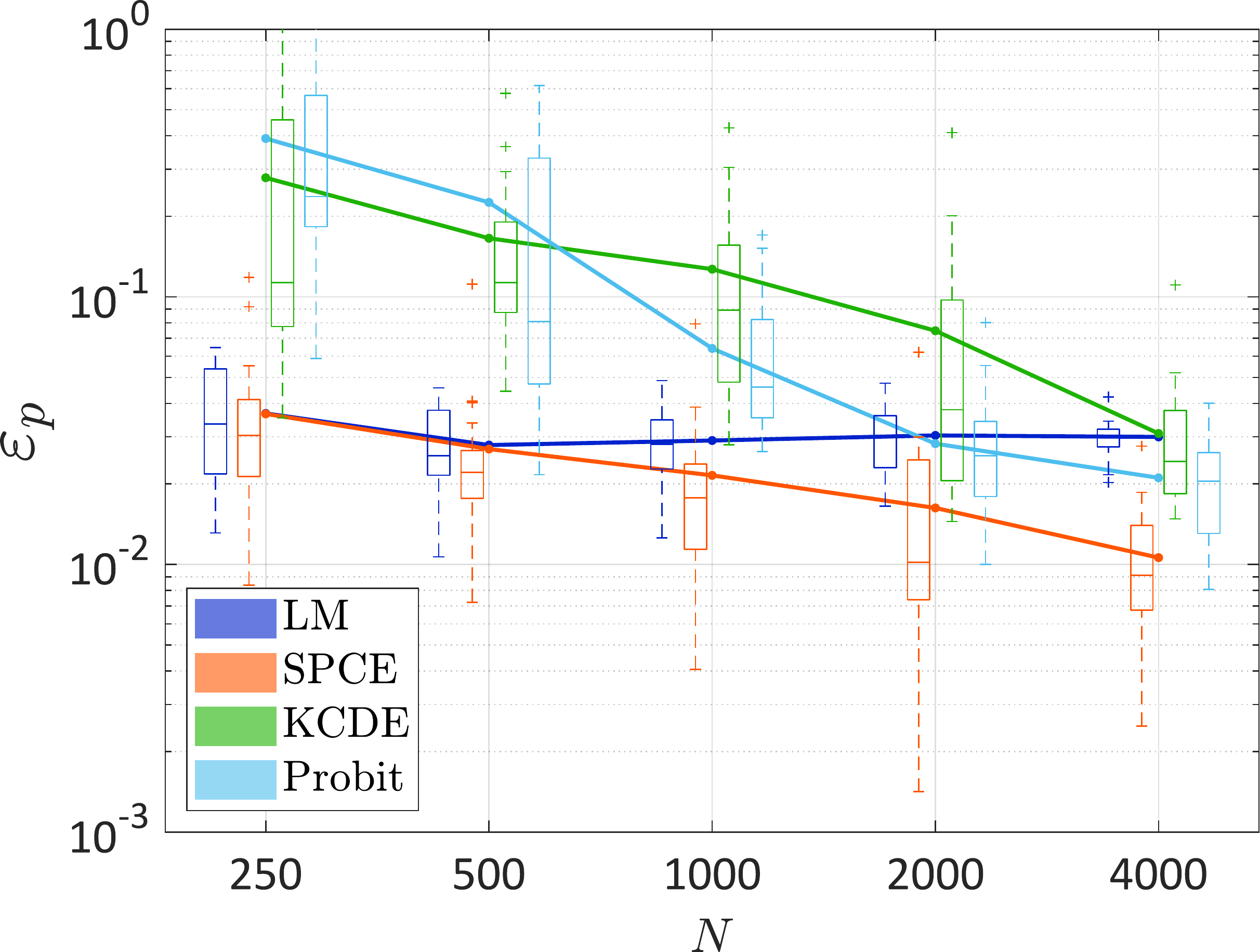}
		\caption{Fragility function for $\delta_0=2.5\%$}		
		\label{fig:Ex2GMPFA2}
	\end{subfigure}
	\caption{Example 2 --- comparison of the convergence among the models in terms of the fragility function. The lines correspond to the average values over 20 repetitions of the full analysis, whereas the box plot summarizes the 20 results.}
	\label{fig:Ex2FA_GMP}
\end{figure}

For fragility function, we select two thresholds $\delta_0 = 0.7\%$ and $\delta_0 = 2.5\%$ which are typically used to characterize light and moderate damages for steel frames \cite{FEMA2000}. The relative mean-squared errors for estimating the associated two fragility functions are reported in \Cref{fig:Ex2FA_GMP}. For the small threshold of $\delta_0 = 0.7\%$, the results are similar to the distribution estimation in \Cref{fig:Ex2WS_GMP}. Specifically: first, the linear model yields the best estimates of the fragility function when small data sets of $N = 250$ are considered, but the errors 
get stagnant with more data; second, KCDE is too flexible to estimate robustly the fragility function due to its nonparametric feature; third, SPCE performs similarly to the linear model for $N=500$ but outperforms all the other models for $N\geq 1{,}000$. Unlike the first example (\Cref{fig:Ex1GMPFA1}), the errors of the probit model are not comparable to these of SPCE but between the linear model and KCDE. For the high threshold of $\delta_0 = 2.5\%$, SPCE is the best model for all values of $N$. The simplicity of the linear model leads to a significant irreducible bias. In contrast, SPCE, KCDE, and the probit model all demonstrate a clear decay of the errors. The kernel estimator has a large spread of errors but a slow convergence of the average value. The probit model performs poorly for $N\leq 500$ because the model ignores the precise values of the EDP and only a few data points exceed the threshold (ca. $1.8\%$ in the data set). In summary, SPCE generally provides more accurate estimates of the fragility functions than the other models.

\begin{figure}[htbp!]
	\centering
	\begin{subfigure}{.48\linewidth}
		\centering
		\includegraphics[height=0.78\linewidth, keepaspectratio]{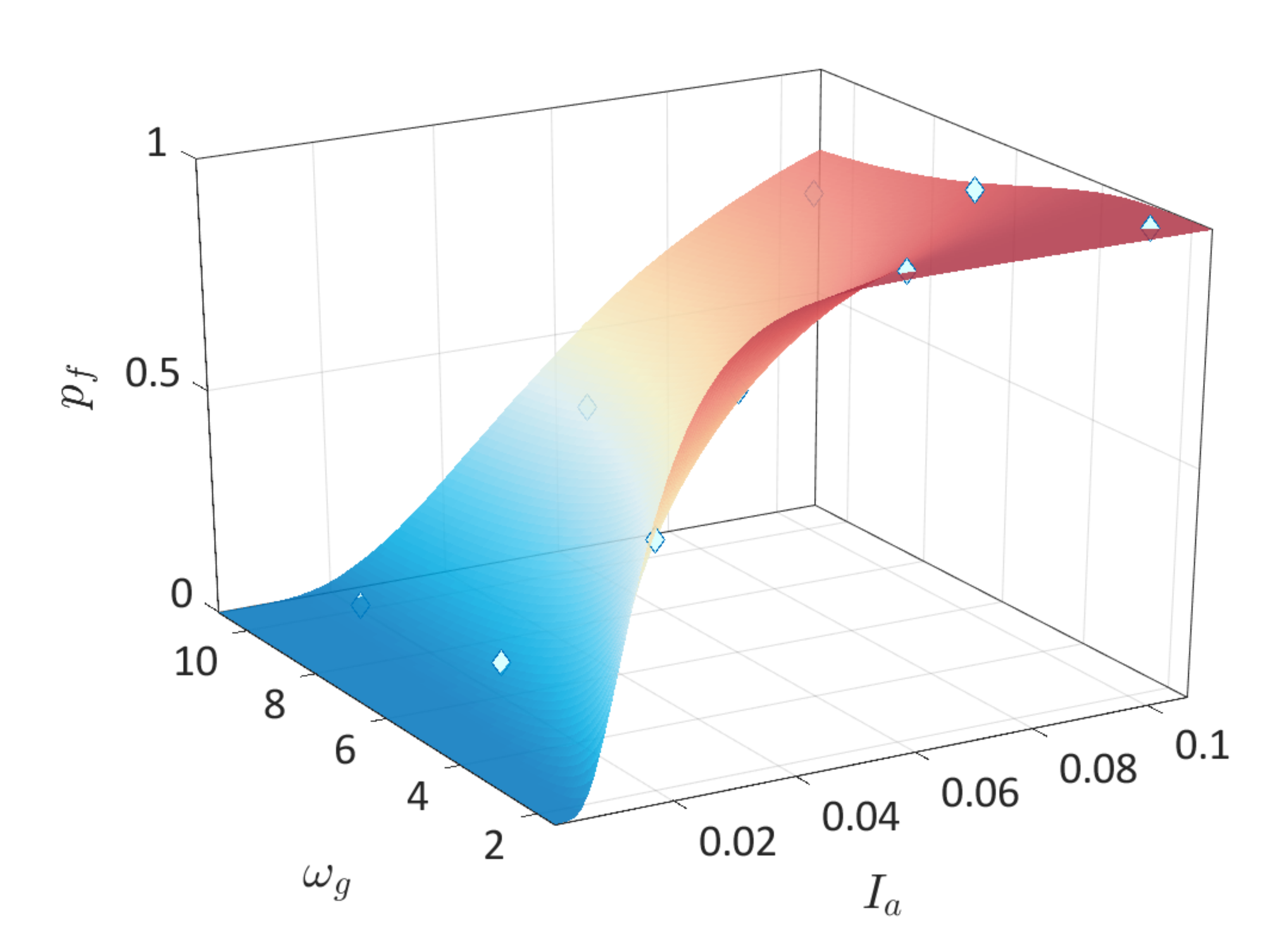}
		\caption{Fragility function for $\delta_0=0.7\%$}
		\label{fig:Ex2FA_surf1}
	\end{subfigure}
	\hfill
	\begin{subfigure}{.48\linewidth}
		\centering
		\includegraphics[height=0.78\linewidth, keepaspectratio]{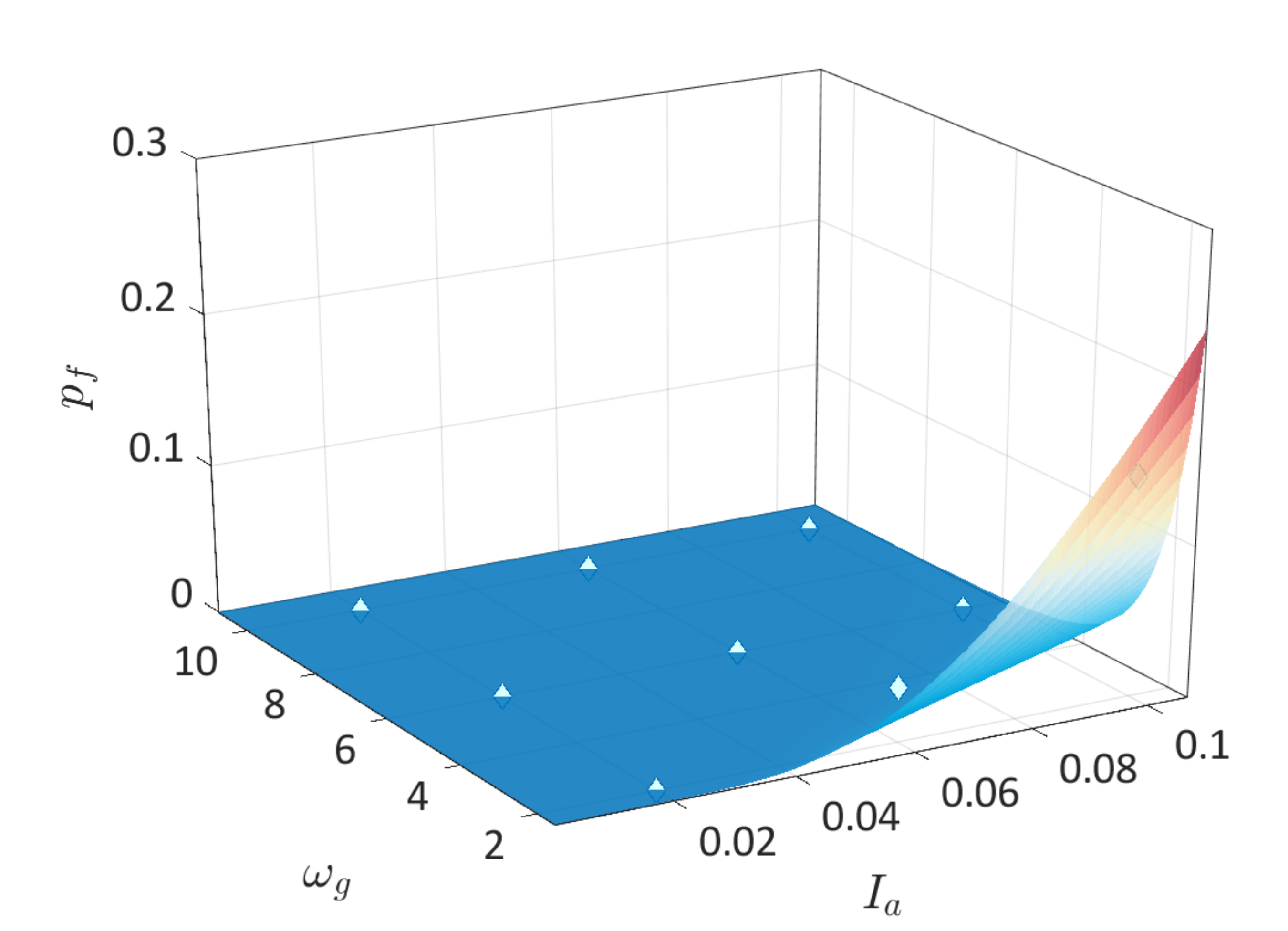}
		\caption{Fragility function for $\delta_0=2.5\%$}		
		\label{fig:Ex2FA_surf2}
	\end{subfigure}
	\caption{Example 2 --- Fragility function in the $I_a-\omega_g$ plan of a SPCE built on $1{,}000$ samples. The diamond points correspond to the reference value computed from $250$ replications.}
	\label{fig:Ex2FA_surf}
\end{figure}

In this example, we plot the two fragility functions in the $I_a-\omega_g$ plan of an SPCE built upon $1{,}000$ model evaluations in \Cref{fig:Ex2FA_surf}. The plotted fragility models are obtained by averaging out the functions with respect to $t_{\text{mid}}$ and $D_{5-95}$. Specifically, we obtain the ``cross-section'' fragility model conditional to each $\{t_{\text{mid}}$, $D_{5-95}\}$ sample, and then, we compute the average fragility model.

We choose $I_a-\omega_g$ which are the most important parameters of the fragility functions according to a sensitivity analysis. This outcome is in line with the results reported in \cite{Abbiati2021}. 
As a comparison, we run the simulator for a validation set of nine points obtained by the Cartesian product of $I_a\in \acc{0.02,0.06,0.1}$ and $\omega_g \in \acc{2,6,10}$. The reference failure probability associated with each validation point is computed by $250$ replications (i.e., a total number of $2{,}250$ simulations for validation). As seen in \Cref{fig:Ex2FA_surf}, the diamonds representing the reference points lie fairly well on the estimated fragility surface. More precisely, the average absolute error of SPCE (averaged over the 9 validation points) is $2.7\%$ for $\delta_0=0.7\%$ and $0.7\%$ for $\delta_0=2.5\%$. In this case, we observe that the dominant variable is the Arias intensity $I_a$ (also confirmed by the sensitivity analysis). This was expected, given the broad-band nature of the excitation, which ``spread'' the energy content among the full range of frequencies. 

\subsection{Discussion}
\noindent
The models considered in this paper were constructed on data without replications. For replication-based approaches \cite{Gidaris2015,Abbiati2021}, a typical number of $\mathcal{O}(10^2)$ replications are used. Following such a strategy, the amount of points exploring the input space would significantly reduce to only $\mathcal{O}(10)$ (as the total number of simulations varies in $\acc{250;500;1{,}000;2{,}000;4{,}000}$). This does not allow for good coverage of the input space, especially when the failure occurs with a higher probability at the tail of the input distribution. Moreover, using replications in the estimation of conditional distributions of a parametric model is also not optimal, as shown in Zhu and Sudret \cite{ZhuSIAMUQ2021}.

Our numerical results demonstrate that SPCE is accurate for estimating both the conditional distribution and the fragility functions for different thresholds. Therefore, SPCE provides a good balance between the model flexibility and limited data. The linear model performs usually well for small values of $N$ but cannot correctly approximate fragility functions with large thresholds. Due to its restrictive assumptions, the linear model cannot be further improved by using more data. Surprisingly, the kernel estimator is almost always the worst model despite that the bandwidth selection procedure is designed for CDF estimation. The probit model directly estimates the fragility function and has a rather low accuracy compared to the other models.

\section{Additional post-processing}
\subsection{CCDF of the EDP}
As the conditional distribution is available from SPCE, one can aggregate the uncertainties in $\ve{X}$ and evaluate the overall risks by uncertainty propagation. As an example, we can compute the complementary cumulative distribution function (CCDF) defined by $\mathbb{P}(Y\geq \delta)$ of the EDP by resampling $Y$ from SPCE. This represents the unconditioned exceeding probability of the EDP as a function of $\delta$. 

\begin{figure}[htbp!]
	\centering
	\begin{subfigure}{.48\linewidth}
		\centering
		\includegraphics[height=0.68\linewidth, keepaspectratio]{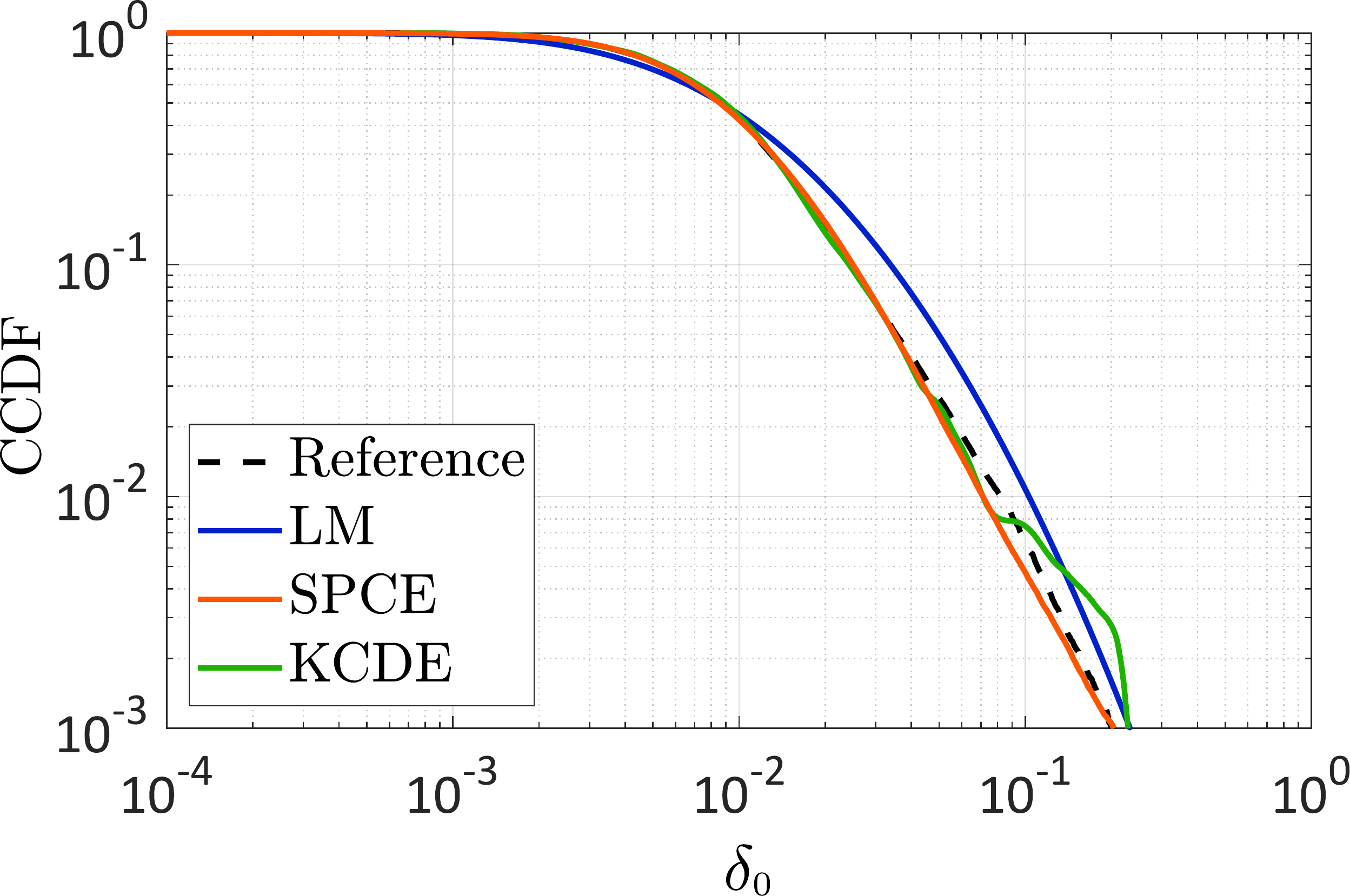}
		\caption{Example 1 (3-DOF system) --- CCDF of the maximum interstory drift}
		\label{fig:Ex1ccdf}
	\end{subfigure}
	\hfill
	\begin{subfigure}{.48\linewidth}
		\centering
		\includegraphics[height=0.68\linewidth, keepaspectratio]{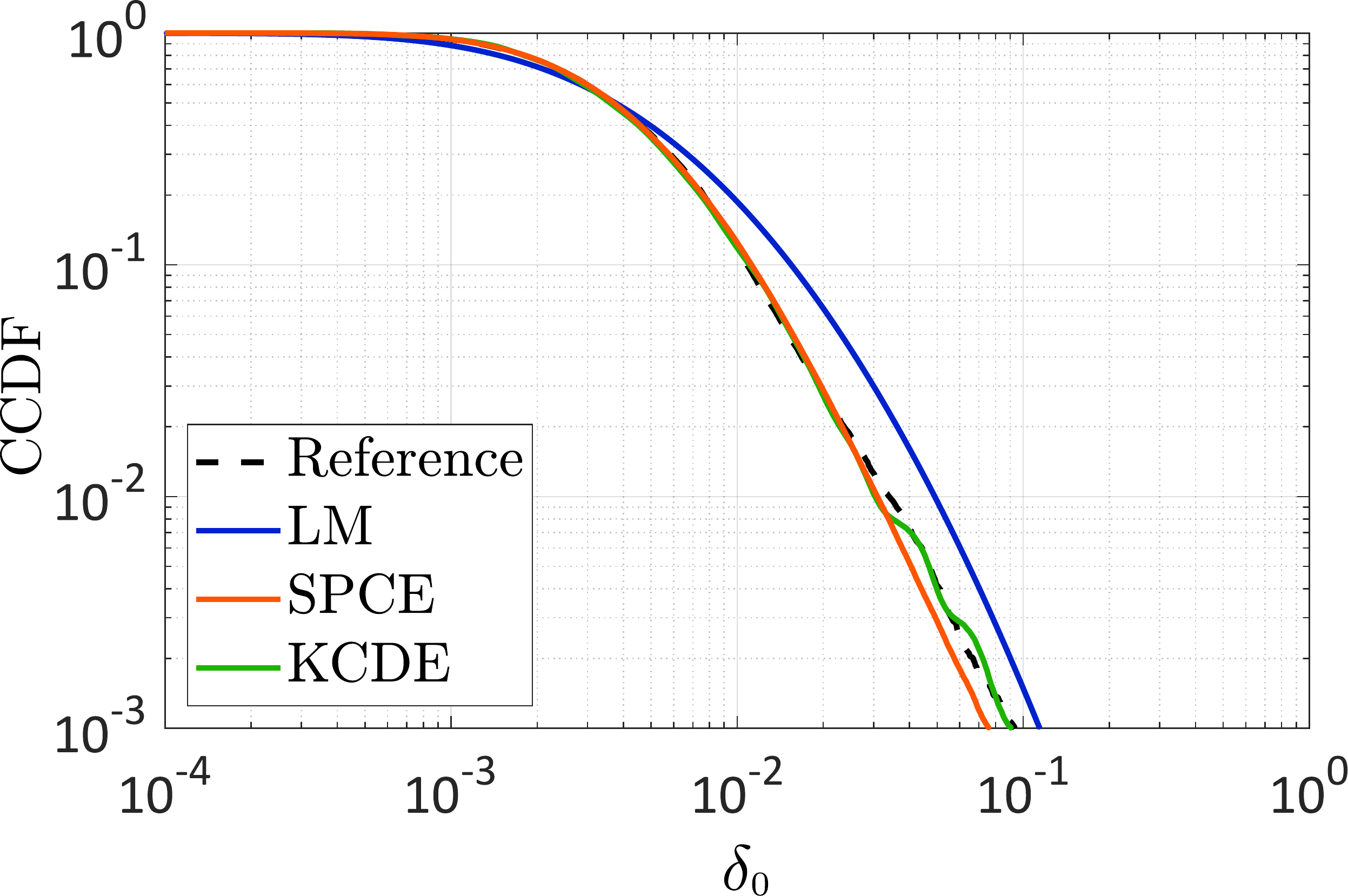}
		\caption{Example 2 (OpenSees model) --- CCDF of the maximum interstory drift ratio}		
		\label{fig:Ex2ccdf}
	\end{subfigure}
	\caption{Comparisons of the CCDF estimation (the models are built on $N=1{,}000$).}
	\label{fig:CCDF_GMP}
\end{figure}

In \Cref{fig:CCDF_GMP}, we plot the CCDFs of the two examples estimated by LM, SPCE, and KCDE, as the probit model does not allow resampling the EDP. The reference curves are the empirical CCDFs using all the available samples ($10^5$ for the first example and $50{,}000$ for the second example). The surrogate models are built on $1{,}000$ simulations. 

We observe that the linear model exhibits a systematic gap at the tail: it overestimates the exceeding probabilities for relatively large values of $\delta$, which cannot be reduced by increasing $N$. The CCDFs obtained from the kernel estimator are generally more accurate than the linear model but are unstable for big values of $\delta$ in \Cref{fig:Ex1ccdf}. SPCE achieves a high accuracy in \Cref{fig:Ex1ccdf} but has a slight discrepancy at the tail in \Cref{fig:Ex2ccdf}) which, according to the numerical investigation, can be efficiently reduced by using more data. 

\subsection{Classical fragility curves}
With the data generated for estimating the distribution of EDP conditioned on the ground motion parameters, we can also compute the fragility curves with respect to a classical IM, such as peak ground acceleration (PGA) or spectral acceleration at the fundamental frequency (SA). More precisely, we first extract the values of the selected IM from the synthetic seismograms and then apply the proposed method to estimate conditional distributions which, by post-processing, gives the fragility curves. As an illustration, we choose a data set of size $1{,}000$ to estimate the fragility curves for each of the examples in \Cref{sec:ex1,sec:ex2}.

\begin{figure}[htbp!]
	\centering
	\begin{subfigure}{.48\linewidth}
		\centering
		\includegraphics[height=0.75\linewidth, keepaspectratio]{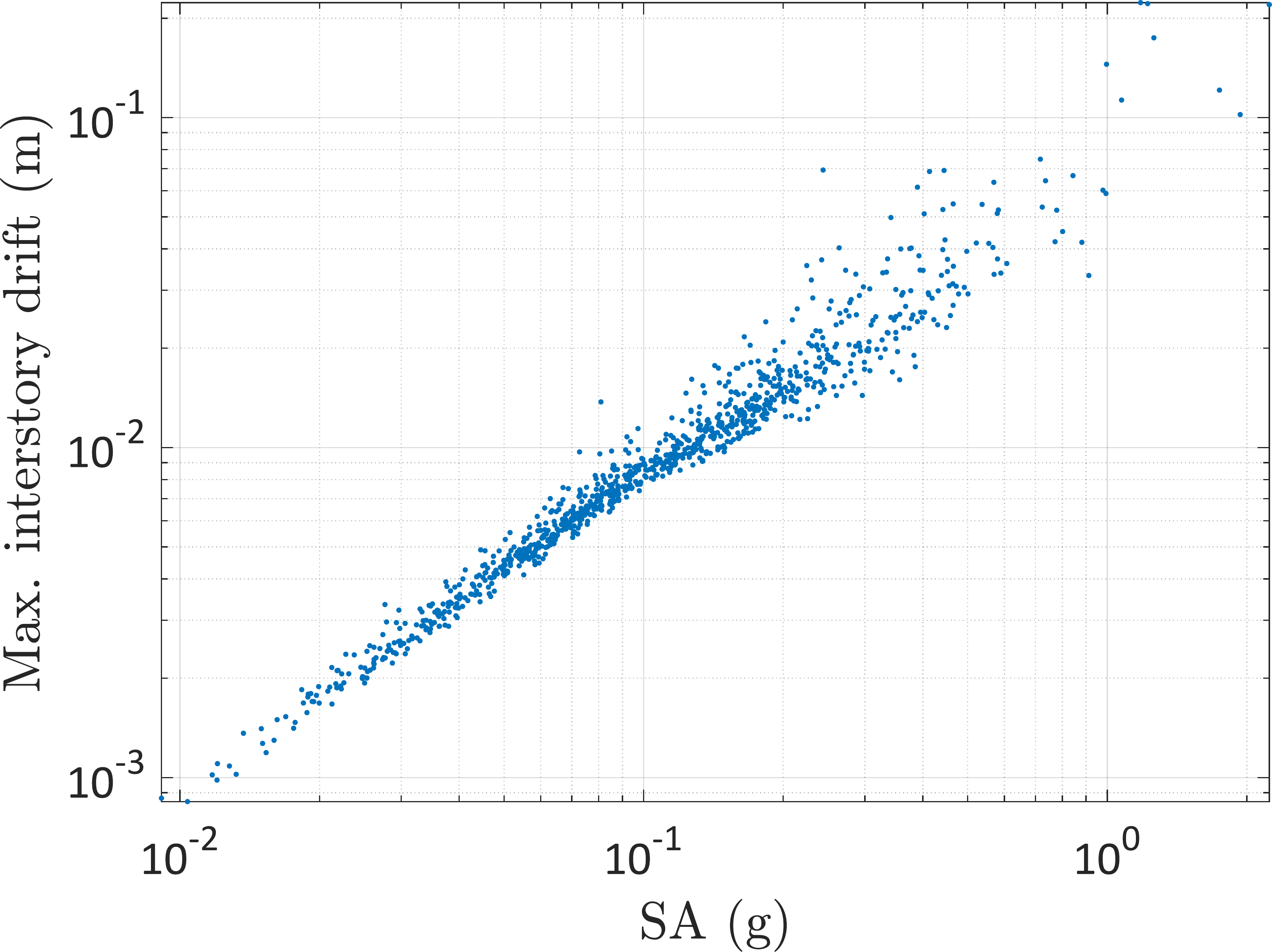}
		\caption{Training samples of size $N=1{,}000$}
		\label{fig:Ex1Samples}
	\end{subfigure}
	\hfill
	\begin{subfigure}{.48\linewidth}
		\centering
		\includegraphics[height=0.75\linewidth, keepaspectratio]{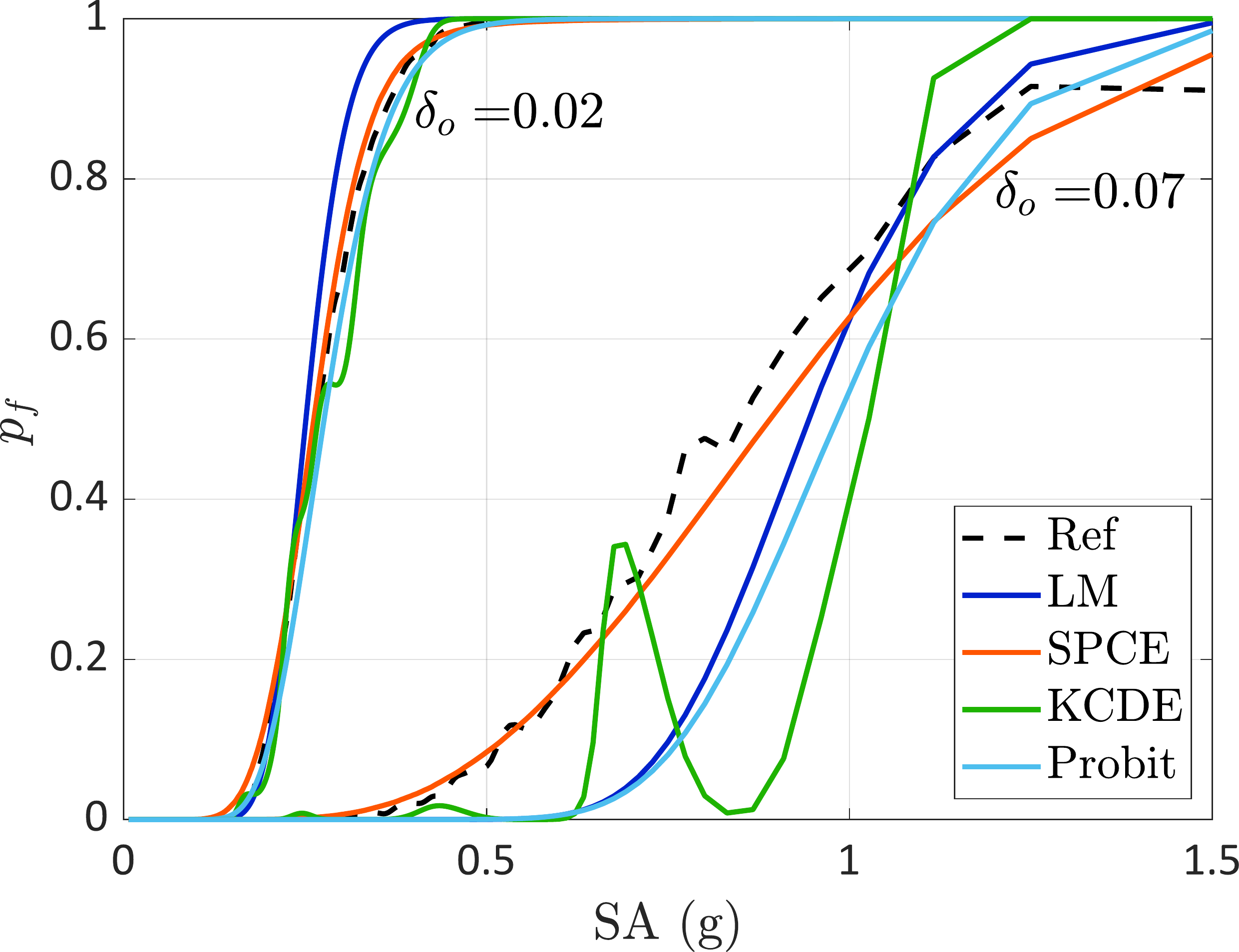}
		\caption{Comparison of fragility curve estimation}	
		\label{fig:Ex1FC}
	\end{subfigure}
	\caption{Example 1 --- fragility curves using spectral acceleration as IM.}
	\label{fig:Ex1FragCurves}
\end{figure}

For the first example (the 3-DOF system), we select the spectral acceleration (SA) as IM. More specifically, SA corresponds to the spectral acceleration for a single-degree-of-freedom system with a period equal to the fundamental period of the structural and viscous damping ratio equal to 2\%. \Cref{fig:Ex1Samples} shows the scatter plot of the $1{,}000$ data points. We observe that the data have a strong heteroskedasticity (in the log-log scale) reflecting a typical nonlinear structural behavior. 

\Cref{fig:Ex1FC} summarizes the fragility curves estimated by the different models (constructed on the data illustrated in \Cref{fig:Ex1Samples}) for $\delta_0 = 0.02$~m and $\delta_0 = 0.07$~m. The reference fragility curves are computed by applying the kernel estimator to all the available data (i.e., $10^5$). Due to the heteroskedastic effect and the possibly non-Gaussian shape of the conditional distribution, the linear model has a significant gap to the reference, in particular for $\delta_0 = 0.07$. KCDE produces an irregular fragility curve for $\delta_0 = 0.07$. The reason is that most of the data are in the region where the intrinsic variability is not significant, which leads to a small value of the selected bandwidth. This results in a large variance of the estimation in the region where the data are sparse, as KCDE is a local estimator. The probit model is quite accurate for $\delta_0 = 0.02$, but it yields an unstable estimate of the fragility curve for $\delta_0 = 0.07$. This is because only a few points ($9$ out of $1{,}000$) lead to exceedance. Finally, SPCE built on the data in \Cref{fig:Ex1FC} approximates the fragility curves with high accuracy.

\begin{figure}[htbp!]
	\centering
	\begin{subfigure}{.48\linewidth}
		\centering
		\includegraphics[height=0.75\linewidth, keepaspectratio]{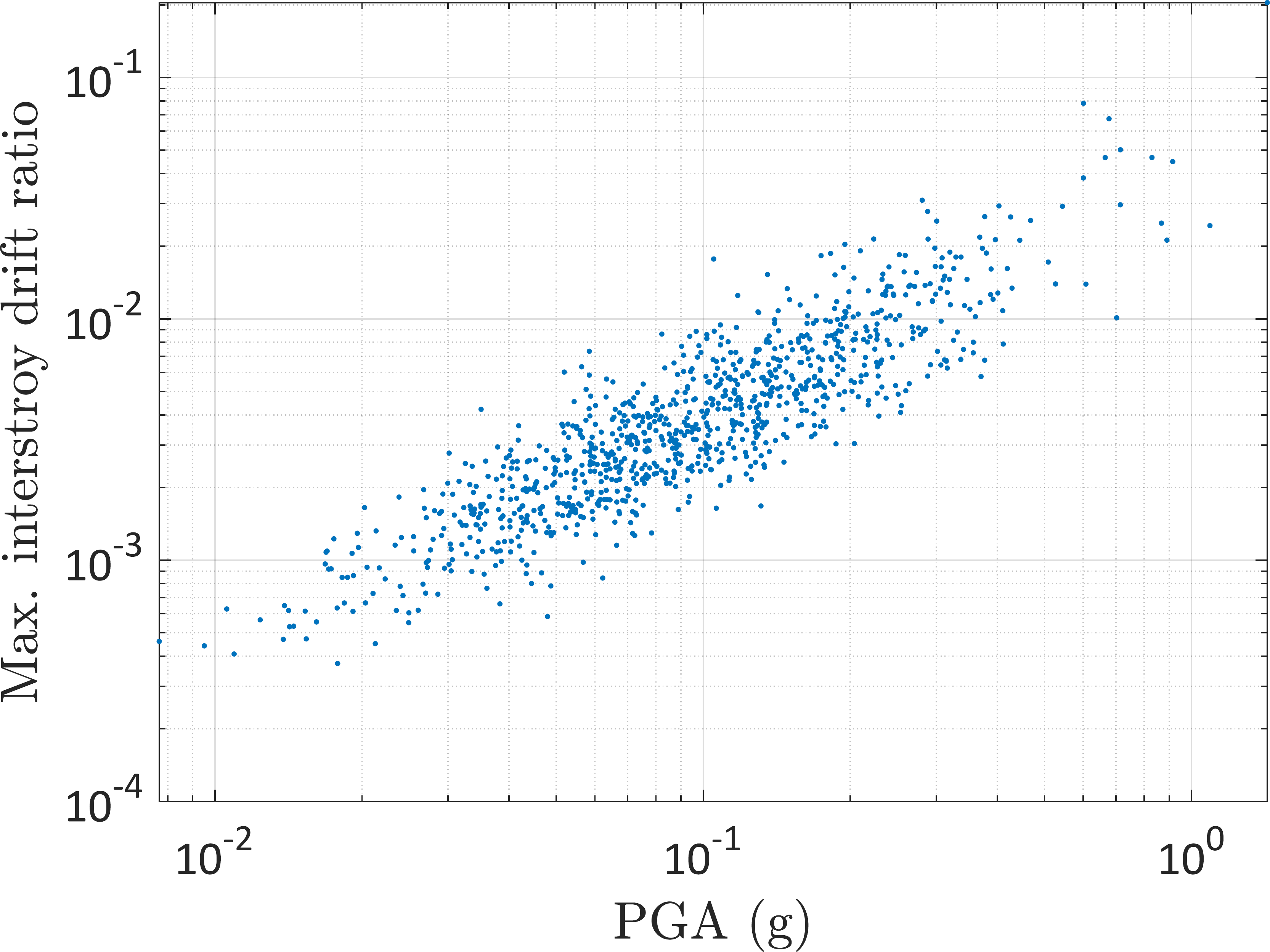}
		\caption{Training samples of size $N=1{,}000$}
		\label{fig:Ex2Samples}
	\end{subfigure}
	\hfill
	\begin{subfigure}{.48\linewidth}
		\centering
		\includegraphics[height=0.75\linewidth, keepaspectratio]{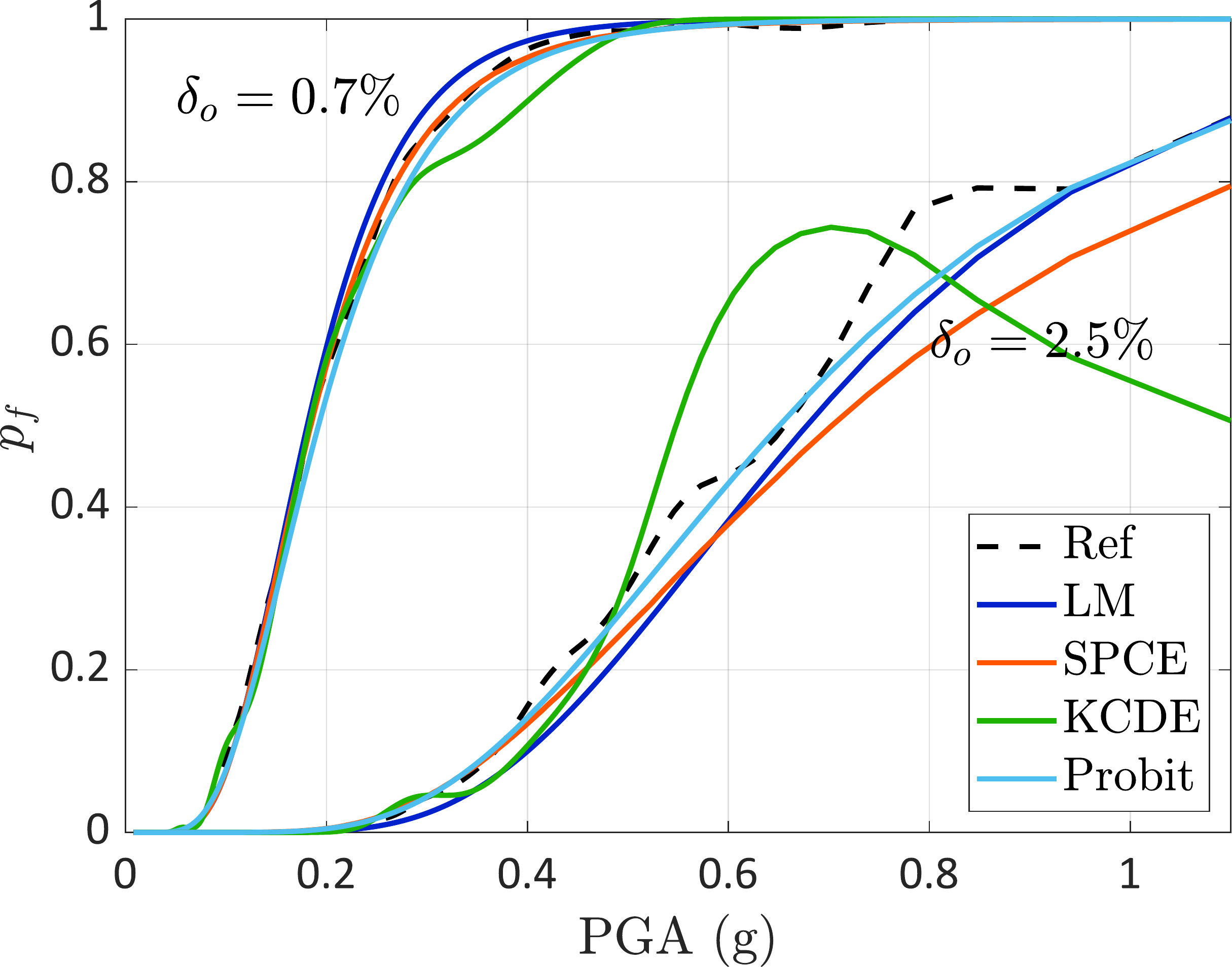}
		\caption{Comparison of fragility curve estimation}	
		\label{fig:Ex2FC}
	\end{subfigure}
	\caption{Example 2 --- fragility curves using peak ground acceleration as IM.}
	\label{fig:Ex2FragCurves}
\end{figure}

For the second example (OpenSees model), we use the PGA as intensity measure. \Cref{fig:Ex2Samples} shows the scatter plot of a data set of size $N = 1{,}000$. We observe that the PGA is a less relevant IM than the SA as the relationship PGA-EDP shows a much larger variability. Therefore, the derived fragility curves are less informative than the previous ones. The PGA-EDP relationship is close to linear with a homoscedastic noise (in the log-log scale). Therefore, the linear model is able to approximate well the fragility curves with relatively small biases. Unlike \Cref{fig:Ex1FC}, the kernel estimator yields smooth predictions, but the fragility curve shows a non-increasing behavior for $\delta_0 = 2.5\%$. The probit model and SPCE provide the most accurate estimate for the fragility curve of $\delta_0 = 0.7\%$. However, in this case, SPCE underestimates the exceeding probabilities associated with the threshold of $\delta_0 = 2.5\%$ for very large values of PGA.
This is because most of the data are in the region where PGA is small and the structure does not fail with very high probabilities: the 95\% and 99\% quantiles of PGA are $0.351 g$ and $0.605 g$, and the associated reference exceeding probabilities are $0.079$ and $0.4427$, respectively.
The SPCE is a flexible model developed to estimate the overall conditional distribution (with respect to the probability distribution of the IM), but not designed to fit directly the tail of the distribution. As a consequence, in specific cases, it may suffer of over-fitting and lack of robust extrapolation behavior for extreme quantiles. In this case, the problem is exacerbated by the relatively large variability between PGA and EDP, which makes difficult the estimation of the tail of the distribution.

The lack of failure data for large damage thresholds is a well-known problem in fragility analysis. In the classical framework for fragility computation based on real ground motions, this problem is overcome by scaling the ground motions and fitting procedures based on censored data \cite{Baker2015}. In the context of stochastic simulation, scaling is not recommended \cite{Grigoriu2011}. A promising future research line is to develop an importance sampling scheme to simulate extreme events from the SGMM model and fill adaptively the EDP intervals of interest. Observe that the presented SPCE approach is orthogonal to this research line and can be easily adapted and applied once the adaptive importance density scheme is developed.

\section{Conclusions}
\label{sec:conclusion}
\noindent
In this paper, we propose methods to efficiently perform fragility analysis based on artificial ground motions, following the recent development in Abbiati et al. \cite{Abbiati2021}. We characterize the ground motion model by a few engineering-meaningful parameters that are calibrated from seismic records and modeled by random variables. Combing this model with the dynamical analysis of structures, we obtain a stochastic simulator: for a given set of ground motion parameters, the engineering demand parameter that characterizes the structural damage is random. Because of this non-deterministic relation, classical surrogate models cannot be used to represent the simulator. 

Some methods that have been developed for estimating classical fragility curves can be extended and applied by regarding the ground motion parameters as multiple intensity measures. To have a reliable model without introducing restrictive assumptions, we propose using the recently developed stochastic surrogate model called \emph{stochastic polynomial chaos expansion} to emulate the conditional distribution. This model introduced an artificial latent variable and a noise variable to reproduce the stochastic behavior of the earthquake simulation. 

The performance of the proposed method is illustrated by two numerical examples: a three-degree-of-freedom system and a 3-story steel frame (modeled in OpenSees). For the conditional distribution estimation, SPCE is compared with the linear model and a state-of-the-art kernel conditional distribution estimator. Using an appropriate error measure defined in \cref{eq:Rlevel1} to assess the accuracy, we observe that the linear model reaches its performance limit for only $N = 250$ simulations because of its simplicity. The kernel estimator is too flexible to have a stable estimate as a consequence of its nonparametric feature. In contrast, SPCE demonstrates a steep decay of the errors and yields the best approximation for $N\geq 1{,}000$. 

For the fragility function, we include the probit model in the comparison. The results show that SPCE prevails over the other models, especially for higher thresholds. In addition, SPCE can be used to propagate the uncertainties in the ground motion parameters to evaluate the overall risks. By resampling the model, SPCE can accurately estimate the complementary cumulative distribution function with limited data even at the tail. Furthermore, one can also apply the method to estimate the fragility curves with respect to classical intensity measures, i.e., PGA and SA.

In this paper, we used a simplified ground motion model based on the Kanai--Tajimi power spectral density, which is a common choice in earthquake engineering to model broad-band far-field ground motions. The associated numerical examples are coherent with this choice, and they are used to illustrate the performance of the novel surrogate model for fragility analysis. Nonetheless, the applicability of the novel surrogate model is not bounded to such a specific ground motion model. For more sophisticated models, one can always follow the proposed framework to build the surrogate model, and the application of the surrogate model to more complex case studies is ongoing. In addition, when many uncertain input parameters are involved, a large amount of data may be necessary to achieve an accurate estimate. To this end, we are exploring new methods to build sparse stochastic polynomial chaos expansions. Alternatively, one can follow Abbiati et al. \cite{Abbiati2021} to select and keep only the most influential input parameters by performing a global sensitivity analysis. In this case, sparse quantile regressions with polynomial chaos expansions remain to be investigated to cope with the framework without replications.

SPCE can generally produce accurate estimates of the fragility curves. However, the data are mostly in the safe region for a high threshold because of the sampling procedure. Thus, extrapolating SPCE for extreme quantiles of the intensity measure with limited data is not reliable. To cope with small exceeding probabilities, adaptive design strategies remain to be explored. The simulation scheme will not simply sample the distribution of the ground motion parameters but adaptively select the samples in the region where the structure is prone to fail to improve the predictive quality of the surrogate model \cite{Echard2011,MarelliSS2018}. This will also benefit the estimation of fragility curves with classical IM. 

We underline that the proposed method can be extended by considering uncertain parameters in the structural properties to tackle a larger set of problems. It can also be applied to model other probabilistic components in PBEE such as relating decision variables (e.g., monetary loss) to structural damage and the damage state to EDP. With these models representing the conditional distributions, one can evaluate the exceeding probability function of the decision variables by resampling (similar to the calculation of the CCDF in \Cref{fig:CCDF_GMP}). Studies in this direction are currently under investigation. 

\section*{Acknowledgment}
This paper is a part of the project ``Surrogate Modeling for Stochastic Simulators (SAMOS)'' funded by the Swiss National Science Foundation (Grant \#200021\_175524), whose support is gratefully acknowledged.

\bibliography{Journal.bib}

\end{document}